\documentclass[11pt]{article}
\usepackage{latexsym}
\usepackage{amssymb}
\usepackage[dvips]{graphicx}
\usepackage{epsfig}
\usepackage{rotating}
\usepackage{amsfonts}
\usepackage{amsmath}
\begin{document}

\begin{titlepage}
\begin{flushright}
{\  }\\
{\ }\\
\end{flushright}

\vspace{10pt}

\begin{center}

{\Large\bf Towards the Quantum Geometry}\\

\vspace{5pt}

{\Large\bf of Saturated Quantum Uncertainty Relations:}\\

%

\vspace{5pt}

{\Large\bf The Case of the $(Q,P)$ Heisenberg Observables}

\vspace{40pt}

Jan Govaerts$^{a,b,c,}$\footnote{Fellow of the Stellenbosch Institute for Advanced Study (STIAS),
Stellenbosh, Republic of South Africa}

\vspace{10pt}

$^{a}${\sl Centre for Cosmology, Particle Physics and Phenomenology (CP3),\\
Institut de Recherche en Math\'ematique et Physique (IRMP),\\
Universit\'e catholique de Louvain (U.C.L.),\\
2, Chemin du Cyclotron, B-1348 Louvain-la-Neuve, Belgium}\\
E-mail: {\em Jan.Govaerts@uclouvain.be}

\vspace{10pt}

$^{b}${\sl National Institute for Theoretical Physics (NITheP),\\
Stellenbosch 7600, Republic of South Africa}

\vspace{10pt}

$^{c}${\sl International Chair in Mathematical Physics and Applications (ICMPA--UNESCO Chair),\\
University of Abomey--Calavi, 072 B. P. 50, Cotonou, Republic of Benin}

\vspace{20pt}

\begin{center}
{\it It is a pleasure to dedicate this article to M. Norbert Hounkonnou\\
on the occasion of his sixtieth birthday}
\end{center}


\vspace{10pt}

\begin{abstract}
\noindent
This contribution to the present Workshop Proceedings outlines a general programme for identifying
geometric structures---out of which to possibly recover quantum dynamics as well---associated
to the manifold in Hilbert space of the quantum states that saturate the Schr\"odinger-Robertson
uncertainty relation associated to a specific set of quantum obser\-va\-bles which characterise a
given quantum system and its dynamics. The first step in such an exploration is addressed herein
in the case of the observables $Q$ and $P$ of the Heisenberg algebra for a single degree of freedom
system. The corresponding saturating states are the well known general squeezed states, whose properties
are reviewed and discussed in detail together with some original results, in preparation of a study
deferred to a separated analysis  of their quantum geometry and of the corresponding path integral
representation over such states.

\end{abstract}

\vspace{40pt}

\noindent
To appear in ``{\sl Mathematical Structures and Applications}", Proceedings of the International Workshop
on Contemporary Problems in Mathematical Physics
(COPROMAPH),\\
October 2016 (Cotonou, Republic of Benin), eds. T. Diagana and B. Toni (Springer, 2018).

\end{center}

\end{titlepage}

\setcounter{footnote}{0}

\section{Introduction}
\label{Intro}

Historically, Heisenberg's uncertainty principle\cite{Heis}
has proved to be pivotal in the emergence of quantum mechanics
as the conceptual paradigm for physics at the smallest distance scales. Still to this day
the uncertainty principle remains a reliable guide in the exploration and the understanding
of the physical consequences of the foundational principles of quantum dynamics.

In its original formulation, Heisenberg suggested that measurements of a quantum particle's
(configuration space) coordinate, $q$, and (conjugate) momentum, $p$, are intrinsically limited
in their precision in a way such that
\begin{equation}
\Delta q\,\Delta p \gtrsim h ,\qquad h=2\pi\hbar,\quad
\hbar=\frac{h}{2\pi}\simeq 6.626\times 10^{-34}\ {\rm J}\cdot{\rm s},
\end{equation}
$\hbar$ being the reduced Planck constant.
Soon thereafter, Schr\"odinger\cite{Schro} as well as Robertson\cite{Robert}
made this statement both more precise and more general for any given pair of self-adjoint, or at least hermitian
quantum observables $A$ and $B$, in the form of the Schr\"odinger-Robertson\footnote{Robertson extended
this statement to an arbitrary number of observables in terms of the determinant
of their covariance matrix of bi-correlations.} uncertainty relation (SR-UR),
\begin{equation}
\left(\Delta A\right)^2\left(\Delta B\right)^2\,\ge\,\frac{1}{4}\langle(-i)[A,B]\rangle^2\,+\,
\frac{1}{4}\langle\left\{A-\langle A\rangle,B-\langle B\rangle\right\}\rangle^2,
\label{eq:SR-UR1}
\end{equation}
where as usual $\left(\Delta A\right)^2=\langle(A-\langle A\rangle)^2\rangle$ and
$\left(\Delta B\right)^2=\langle(B-\langle B\rangle)^2\rangle$, while $\langle{\cal O}\rangle$
denotes the normalised expectation value of any quantum operator ${\cal O}$ given
an arbitrary (normalisable) quantum state (see Appendix A for notations and
a derivation of the SR-UR). As a by-product one thus also obtains
the less tight (but better known, and generalised) Heisenberg uncertainty relation (H-UR),
\begin{equation}
\left(\Delta A\right)\,\left(\Delta B\right)\,\ge\,\frac{1}{2}|\langle(-i)[A,B]\rangle|.
\label{eq:H-UR1}
\end{equation}
In the case of the Heisenberg algebra, namely $[Q,P]=i\hbar\,\mathbb{I}$,
indeed this becomes $\Delta q\,\Delta p\ge \hbar/2$.

In the classical limit $\hbar\rightarrow 0$, both terms of these inequalities vanish and the latter turn into
strict equalities. The physical world however, is not classical since Planck's constant albeit small as measured
in our macroscopic units, definitely has a finite and non-vanishing value. Yet, in certain regimes of their
Hilbert spaces dynamical quantum systems must display classical behaviour as we experience it through
quantum observables some of which are of a macroscopic character. Indeed, any quantum system is specified through
a set of quantum observables of which the algebra of commutation relations is represented by the Hilbert
space which describes that quantum system and its quantum states.
Given a particular choice of quantum observables and through measurements of the latter,
experiments give access to the quantum states of such a system and enable their manipulation.
If certain regimes of a quantum system display the hallmarks of a classical-like behaviour, certainly these regimes
must correspond to quantum states which are as close as possible to being classical given a set ensemble of
quantum observables characterising that system. In other words classical-like regimes of a quantum system
which is characterised by a collection of quantum observables, need to correspond to quantum states
which saturate as exact equalities the generalised Schr\"odinger-Robertson uncertainty relation related
to that ensemble of quantum observables. Indeed saturated uncertainty relations leave the least room possible
for a genuine quantum dynamical behaviour which would otherwise potentially lead to large differences
in the values taken by the two terms involved in the inequalities expressing such uncertainty relations.

For reasons recalled in Appendix A, in the case of two observables the SR-UR is saturated
by quantum states $|\psi_0\rangle$ which are such that,
\begin{equation}
\left[(A-\langle A\rangle)\,-\,\lambda_0\,(B-\langle B\rangle)\right]\,|\psi_0\rangle = 0,\qquad
\left[A\,-\,\lambda_0\,B\right]\,|\psi_0\rangle =
\left[\langle A\rangle\,-\,\lambda_0\,\langle B\rangle\right]\,|\psi_0\rangle,
\label{eq:rel1}
\end{equation}
where the complex parameter $\lambda_0$ is given by the following combination of expectation values
for the state $|\psi_0\rangle$,
\begin{equation}
\lambda_0=\frac{\langle(B-\langle B\rangle)(A-\langle A\rangle)\rangle}{\left(\Delta B\right)^2}=
\frac{\left(\Delta A\right)^2}{\langle(A-\langle A\rangle)(B-\langle B\rangle)\rangle}.
\end{equation}
Such saturating quantum states are parametrised by collections of continuous parameters,
if only for the expectation values $\langle A\rangle$ and $\langle B\rangle$ as well as the ratio
$\Delta A/\Delta B$, for instance. Indeed, especially when considered in the form of the second relation in (\ref{eq:rel1}),
such states determine classes of quantum coherent-like states (see Refs.\cite{Klauder1,Nieto}
and references therein), which share many of
the remarkable properties of the well known Schr\"odinger canonical coherent states for the Heisenberg algebra.
In particular in order that their expectation values $\langle A\rangle$ and $\langle B\rangle$ retain finite non-vanishing
classical values as $\hbar\rightarrow 0$, it is necessary that the saturating states $|\psi_0\rangle$ meeting
the conditions (\ref{eq:rel1}) involve all possible linearly independent quantum states spanning
the full Hilbert space of the system. Furthermore, usually the linear span of such coherent states encompasses
the full Hilbert space, since they obey a specific overcompleteness relation or resolution of the unit operator,
thereby providing a self-reproducing kernel representation of that Hilbert space\cite{Klauder1,Klauder2}.

In other words, given a set of quantum observables such saturating quantum states for the corresponding collection
of uncertainty relations determine a specific differentiable submanifold of Hilbert space, out of which
the full Hilbert space of the quantum system may {\sl a priori\/} be reconstructed (provided a sufficient
number of quantum observables is considered). In particular quantum amplitudes may then be given a functional
path integral representation over that manifold of coherent states, which involves specific geometrical structures
of that manifold\cite{Klauder2,Klauder3}.
Indeed, very naturally that manifold comes equipped then not only with a (quantum) symplectic
structure\footnote{Because of the sesquilinear properties of the inner product defined over Hilbert space.}
but also with a (quantum) Riemannian metric structure\footnote{Because of the hermitian and
positive definite properties of the inner product defined over Hilbert space.}\cite{Klauder3},
both of these geometric structures
being compatible with one another (and dependent, generally, on Planck's constant).
A quantum geometric representation of the quantum system thus arises out
of its Hilbert space given a choice of its quantum observables and through the associated uncertainty relation.
It may even be that, for instance through the corresponding path integral, the quantum system itself may
be reconstructed out of these geometric structures (provided the original choice of quantum observables
be large enough).

Such an approach connects directly with, and expands on Klauder's general programme of ``Enhanced Quantisation"
having been proposed for many years now (see Ref.\cite{Klauder2} and references therein),
as a path towards a geometrical formulation of genuine quantum
dynamics which shares a number of similarities with other proposals for such geometrical formulations\cite{Pro,Ash}.
For that same reason, the programme as briefly outlined above provides a possible avenue towards
a further understanding of the underpinnings of the AdS/CFT correspondence and the holographic principle,
for instance along lines similar to those having been explored already in Ref.\cite{Kriel}.

While the general programme outlined above, based on saturated uncertainty relations and the geometry
of the associated coherent-like quantum states, is offered
here as a project of possible interest to Professor Norbert Hounkonnou in celebration as well of his sixtieth
birthday and on the occasion of this COPROMAPH Workshop organised in his honour, the present paper only deals
with the construction of the quantum states which saturate the Schr\"odinger-Robertson uncertainty relation
in the case of the Heisenberg algebra for a single quantum degree of freedom, leaving for separate work
a discussion of the ensuing geometric structures.
Besides some results which presumably are original, most of those being presented herein certainly
are available in the literature (see Refs.\cite{Trifonov,Angelow,Brif} and references therein)
even though in a scattered form\footnote{For this reason no attempt is being made
towards a complete list of references to the original literature which relates to many different fields of
quantum physics.}. However this author did not find them discussed
along the lines addressed here, nor could he find them all brought together in one single place, as made available
in the present contribution with the purpose of providing a basis towards a pursuit of the projected programme
aiming at a better understanding of the geometric structures inherent to quantum systems and their dynamics.

Section \ref{Sect2} particularises the discussion to the Heisenberg algebra and identifies the saturating states
for the SR-UR in the configuration space representation of that algebra. A construction in terms of Fock algebras
and their canonical coherent states is then initiated in Section \ref{Sect3}, beginning with a reference Fock algebra
related to an intrinsic physical scale. Section \ref{Sect4} then presents the complete parametrised set of saturating
quantum states, leading to the general class of the well known squeezed coherent states. Further specific results
of interest for these states are then presented in Sections \ref{Sect5} and \ref{Sect6}, to conclude
with some additional comments in the Conclusions. Complementary material of a more pedagogical character
as befits the Proceedings of the present COPROMAPH Workshop, is included in two Appendices.

\section{The Uncertainty Relation for the Heisenberg Algebra}
\label{Sect2}

Given a single degree of freedom system whose configuration space has the topology of the
real line, $q\in\mathbb{R}$, let us consider the corresponding Heisenberg algebra with its
conjugate quantum observables, $Q$ and $P$, such that
\begin{equation}
\left[Q,P\right]=i\hbar\,\mathbb{I},\qquad
Q^\dagger=Q,\quad P^\dagger=P.
\end{equation}
The configuration and momentum space representations of this algebra are well known,
based on the corresponding eigenstate bases, $Q|q\rangle=q|q\rangle$ and $P|p\rangle=p|p\rangle$,
with $q,p\in\mathbb{R}$. Our choices of normalisations and phase conventions for these bases states
are such that\footnote{Hence the states $|q\rangle$, say, are determined up to a $q$-independent
overall global phase factor which remains unspecified, relative to which all other phase factors are then
identified accordingly.}
\begin{equation}
\langle q|q'\rangle=\delta(q-q'),\quad
\langle p|p'\rangle=\delta(p-p'),\quad
\int_{-\infty}^{+\infty}dq\,|q\rangle\langle q|=\mathbb{I}=\int_{-\infty}^{+\infty}dp\,|p\rangle\langle p|,
\end{equation}
\begin{equation}
\langle q|p\rangle=\frac{1}{\sqrt{2\pi\hbar}}e^{\frac{i}{\hbar}qp},\qquad
\langle p|q\rangle=\frac{1}{\sqrt{2\pi\hbar}}e^{-\frac{i}{\hbar}qp}.
\end{equation}

Consider an arbitrary (normalisable) quantum state $|\psi_0\rangle$, which we assume also to have been normalised,
$\langle\psi_0|\psi_0\rangle=1$. In configuration space this state is represented by its wave function,
$\psi_0(q)=\langle q|\psi_0\rangle\in\mathbb{C}$. Let $q_0$ and $p_0$ be its real valued expectation values for the
Heisenberg observables, 
\begin{equation}
q_0=\langle\psi_0|Q|\psi_0\rangle,\qquad
p_0=\langle\psi_0|P|\psi_0\rangle,\qquad
q_0,p_0\in\mathbb{R},
\end{equation}
and introduce the shifted or displaced operators
\begin{equation}
\bar{Q}=Q-q_0,\qquad
\bar{P}=P-p_0,
\end{equation}
which again define a Heisenberg algebra of hermitian (ideally self-adjoint) quantum observables,
$[\bar{Q},\bar{P}]=i\hbar\,\mathbb{I}$, $\bar{Q}^\dagger=\bar{Q}$, $\bar{P}^\dagger=\bar{P}$.
One also has $\left(\Delta Q\right)^2=\langle\psi_0|\bar{Q}^2|\psi_0\rangle$ and
$\left(\Delta P\right)^2=\langle\psi_0|\bar{P}^2|\psi_0\rangle$.

The Schr\"odinger-Robertson uncertainty relation (SR-UR) then reads (see Appendix A),
\begin{equation}
\left(\Delta Q\right)^2\,\left(\Delta P\right)^2\,\ge\,\frac{1}{4}\hbar^2\,+\,
\frac{1}{4}\langle\left\{\bar{Q},\bar{P}\right\}\rangle^2,
\end{equation}
$\left\{\bar{Q},\bar{P}\right\}$ being the anticommutator of $\bar{Q}$ and $\bar{P}$.
As a corollary note that one then also has the looser Heisenberg uncertainty relation (H-UR),
\begin{equation}
\Delta Q\,\Delta P\,\ge\,\frac{1}{2}\hbar.
\end{equation}
However according to the general programme outlined in the Introduction, we are interested in identifying
all quantum states that saturate the SR-UR, but not necessarily the H-UR. Quantum states that saturate
the H-UR are certainly such that $\langle\left\{\bar{Q},\bar{P}\right\}\rangle=0$, namely they cannot possess
any $(Q,P)$ quantum correlation. The ensemble of states that saturate the SR-UR is thus certainly larger
than that which saturates the H-UR. What distinguishes these two sets of states will be made explicit later on.

For reasons recalled in Appendix A, those states which saturate the SR-UR are such that
\begin{equation}
\left[\bar{Q}\,-\,\lambda_0\,\bar{P}\right]|\psi_0\rangle=0,\qquad
\left[\left(Q-q_0\right)\,-\,\lambda_0\,\left(P-p_0\right)\right]\,|\psi_0\rangle=0,
\label{eq:defining1}
\end{equation}
where the complex parameter $\lambda_0$ takes the value,
\begin{equation}
\lambda_0=\frac{1}{\left(\Delta P\right)^2}\left(\frac{1}{2}\langle\left\{\bar{Q},\bar{P}\right\}\rangle\,-\,\frac{1}{2}i\hbar\right)
=\left(\Delta Q\right)^2\frac{1}{\frac{1}{2}\langle\left\{\bar{Q},\bar{P}\right\}\rangle\,+\,\frac{1}{2}i\hbar}.
\end{equation}
The defining equation (\ref{eq:defining1}) of saturating states for the $(Q,P)$ observables of the Heisenberg algebra
is best solved by working in a wave function representation, say in configuration space.
The above condition then reads,
\begin{equation}
\left[\left(q-q_0\right)-\lambda_0\left(-i\hbar\frac{d}{dq} - p_0\right)\right]\,\psi_0(q)=0.
\end{equation}
Clearly its solution is
\begin{equation}
\psi_0(q)=N_0(q_0,p_0,\lambda_0)\ e^{\frac{i}{\hbar}qp_0}\,e^{\frac{i}{2\lambda_0\hbar}(q-q_0)^2},\qquad
\psi^*_0(q)=N^*_0(q_0,p_0,\lambda_0)\ e^{-\frac{i}{\hbar}qp_0}\,e^{-\frac{i}{2\lambda^*_0\hbar}(q-q_0)^2},
\label{eq:wf1}
\end{equation}
where $N_0(q_0,p_0,\lambda_0)$ is a complex valued normalisation factor still to be determined. Requiring the state
$|\psi_0\rangle$ to be normalised to unity implies the following value for the norm of $N_0(q_0,p_0,\lambda_0)$,
\begin{equation}
|N_0(q_0,p_0,\lambda_0)|=\left(2\pi\left(\Delta Q\right)^2\right)^{-1/4}.
\label{eq:N0}
\end{equation}
Its overall phase however, will be determined later on, once further phase conventions will have been specified.
Note well that all quantum states saturating the SR-UR are of this simple form, specified in terms of four independent
real parameters, namely $q_0$, $p_0$, $\Delta Q>0$ (say) and $\langle\left\{\bar{Q},\bar{P}\right\}\rangle$ (in terms
of which $\Delta P>0$ is then also determined since $\left(\Delta Q\right)^2\left(\Delta P\right)^2=(\hbar^2+
\langle\left\{\bar{Q},\bar{P}\right\}\rangle^2)/4$). In the remainder of this paper, we endeavour to understand
the structure of these saturating quantum states from the point of view of coherent states, as indeed the defining
equation (\ref{eq:defining1}) invites us to do.

To conclude, let us also remark that for those saturating states such that in addition
$\langle\left\{\bar{Q},\bar{P}\right\}\rangle=0$, in this particular case which thus saturates the H-UR
rather than the SR-UR we have the following results
(with a choice of phase factor for the wave function which complies with the specifications to be addressed later on),
\begin{equation}
\langle\left\{\bar{Q},\bar{P}\right\}\rangle=0:\qquad
\lambda_0=-\frac{i\hbar}{2\left(\Delta P\right)^2}=-2i\frac{\left(\Delta Q\right)^2}{\hbar},\qquad
\frac{1}{\lambda_0}=\frac{i\hbar}{2\left(\Delta Q\right)^2},
\end{equation}
\begin{equation}
\psi_0(q)=\frac{1}{\left(2\pi\left(\Delta Q\right)^2\right)^{1/4}}\ e^{\frac{i}{\hbar}q p_0}
\,e^{-\frac{1}{4\left(\Delta Q\right)^2}(q-q_0)^2},\qquad
\Delta Q\,\Delta P=\frac{1}{2}\hbar.
\end{equation}
Note well however, that even in this case the value of $\Delta P/\Delta Q=\hbar/(2(\Delta Q)^2)$
is still left as a free real and positive parameter. In the case of the ordinary Schr\"odinger canonical
coherent states, which indeed saturate the H-UR, this latter ratio is implicitly set to a specific value
in terms of physical parameters of the system under consideration.

\section{A Reference Fock Algebra}
\label{Sect3}

{\sl A priori} the quantum observables $Q$ and $P$ possess specific physical dimensions,
of which the product has the physical dimension of $\hbar$.
For the sake of the construction hereafter, let us denote by $\ell_0$ an intrinsic physical scale
which has the same physical dimension as $Q$, so that the physical dimension of $P$ is that of $\hbar/\ell_0$.
For instance we may think of $Q$ as a configuration space coordinate measured in a unit of length,
in which case $\ell_0$ has the dimension of length, hence the notation.
However, note that the physical dimension of $\ell_0$ could be anything, as may be relevant
given the physical system under consideration. Furthermore $\ell_0$ need not correspond to some fundamental physical
scale or constant. The scale $\ell_0$ may well be expressed in terms of fundamental physical constants in combination
with other physical parameters related to the system under consideration. In particular $\ell_0$ may involve
Planck's constant itself, $\hbar$, and thus change value in the classical limit $\hbar\rightarrow 0$
(as is the case for the ordinary harmonic oscillator of mass $m$ and angular frequency $\omega$,
with then the natural choice $\ell_0=\sqrt{\hbar/(m\omega)}$). The purpose of the intrinsic physical scale $\ell_0$
is to introduce a reference quantum Fock algebra, hence the corresponding reference canonical coherent states,
in order to address the quantum content characterised by the defining equation (\ref{eq:defining1})
of the quantum states saturating the SR-UR of the Heisenberg algebra, which is indeed
a condition characteristic of quantum coherent states.

Given the intrinsic physical scale $\ell_0$, let us thus introduce the following reference Fock operators,
\begin{equation}
a=\frac{1}{\sqrt{2}}\left(\frac{Q}{\ell_0}\,+\,i\frac{\ell_0}{\hbar} P\right),\qquad
a^\dagger=\frac{1}{\sqrt{2}}\left(\frac{Q}{\ell_0}\,-\,i\frac{\ell_0}{\hbar} P\right),
\label{eq:Fock-a}
\end{equation}
with the inverse relations for the Heisenberg observables,
\begin{equation}
Q=\frac{1}{\sqrt{2}}\ell_0\left(a+a^\dagger\right),\qquad
P=-\frac{i\hbar}{\ell_0\sqrt{2}}\left(a-a^\dagger\right),
\end{equation}
which indeed generate the corresponding Fock and Heisenberg algebras, respectively,
\begin{equation}
\left[ a, a^\dagger\right]=\mathbb{I},\qquad
\left[Q,P\right]=i\hbar\,\mathbb{I}.
\end{equation}
The associated normalised reference Fock vacuum, $|\Omega_0\rangle$, such that
\begin{equation}
a|\Omega_0\rangle=0,\qquad
\langle\Omega_0|\Omega_0\rangle = 1,
\end{equation}
is chosen with a phase relative to the overall phase implicitly chosen for the position eigenstates $|q\rangle$
such that
\begin{equation}
\langle q|\Omega_0\rangle= \left(\pi\ell^2_0\right)^{-1/4}\,e^{-\frac{1}{2\ell^2_0}q^2}.
\end{equation}
On account of the condition $a|\Omega_0\rangle=0$ to be compared to the defining equation (\ref{eq:defining1}),
it is clear that the reference Fock vacuum $|\Omega_0\rangle$ saturates not only the SR-UR but also
the H-UR with vanishing expectation values for $q_0$, for $p_0$ and for the $(Q,P)$ correlator
$\langle\left\{Q,P\right\}\rangle$, while the values for $\Delta Q$ and $\Delta P$ given by
\begin{equation}
\left(\Delta Q\right)^2=\frac{1}{2}\ell^2_0,\quad
\left(\Delta P\right)^2=\frac{1}{2}\frac{\hbar^2}{\ell^2_0},\quad
\left(\frac{\Delta Q}{\ell_0}\right)^2=\frac{1}{2},\quad
\left(\frac{\ell_0}{\hbar}\Delta P\right)^2=\frac{1}{2},
\end{equation}
are such that,
\begin{equation}
\left(\Delta Q\right)\,\left(\Delta P\right)=\frac{1}{2}\hbar,\quad
\left(\frac{\Delta Q}{\ell_0}\right)\,\left(\frac{\ell_0}{\hbar}\Delta P\right)=\frac{1}{2},\quad
\left(\frac{\Delta Q}{\ell_0}\right)^2\,+\,\left(\frac{\ell_0}{\hbar}\Delta P\right)^2 = 1,
\end{equation}
with in particular thus even the ratio $\Delta P/\Delta Q$ taking a predetermined value,
$\Delta P/\Delta Q=\hbar/\ell^2_0$. As it turns out, all states saturating the SR-UR will
be constructed out of this reference Fock vacuum (thereby also determining the overall phase
of the wave function of these states, $\psi_0(q)$, left unspecified in (\ref{eq:wf1}) and
(\ref{eq:N0}) of Section \ref{Sect2}).

In order to deal with the shifted or displaced observables $\bar{Q}$ and $\bar{P}$ which involve the
expectation values $q_0$ and $p_0$, given the reference Fock algebra (\ref{eq:Fock-a})
let us introduce the following complex quantity,
\begin{equation}
u_0=\frac{1}{\sqrt{2}}\left(\frac{q_0}{\ell_0}+i\frac{\ell_0}{\hbar}p_0\right),\qquad
\bar{u}_0=u^*_0=\frac{1}{\sqrt{2}}\left(\frac{q_0}{\ell_0}-i\frac{\ell_0}{\hbar}p_0\right),
\end{equation}
with the inverse relations,
\begin{equation}
q_0=\frac{1}{\sqrt{2}}\ell_0\left(u_0 + \bar{u}_0\right),\qquad
p_0=-\frac{i\hbar}{\ell_0\sqrt{2}}\left(u_0 - \bar{u}_0\right).
\end{equation}
Correspondingly we have the Fock algebra of the associated shifted or displaced Fock generators,
\begin{equation}
b(u_0)=a-u_0,\qquad
b^\dagger(u_0)=a^\dagger-\bar{u}_0,\qquad
\left[b(u_0),b^\dagger(u_0)\right]=\mathbb{I},
\end{equation}
which are such that,
\begin{equation}
b(u_0)=\frac{1}{\sqrt{2}}\left(\frac{\bar{Q}}{\ell_0}\,+\,i\frac{\ell_0}{\hbar} \bar{P}\right),\qquad
b^\dagger(u_0)=\frac{1}{\sqrt{2}}\left(\frac{\bar{Q}}{\ell_0}\,-\,i\frac{\ell_0}{\hbar} \bar{P}\right),
\end{equation}
as well as,
\begin{equation}
\bar{Q}=\frac{1}{\sqrt{2}}\ell_0\left(b(u_0)+b^\dagger(u_0)\right),\qquad
\bar{P}=-\frac{i\hbar}{\ell_0\sqrt{2}}\left(b(u_0)-b^\dagger(u_0)\right),\qquad
\left[\bar{Q},\bar{P}\right]=i\hbar\,\mathbb{I}.
\end{equation}
The correspondence between the displaced Fock algebra and the reference one is best understood
by considering the displacement operator\cite{Klauder1} defined\footnote{All Baker-Campbell-Hausdorff (BCH) formulae
necessary for this paper are discussed in Appendix B.} in terms of the parameters $u_0$ or $(q_0,p_0)$,
\begin{equation}
D(q_0,p_0)\equiv D(u_0)\equiv e^{u_0a^\dagger - \bar{u}_0 a}=
e^{-\frac{1}{2}|u_0|^2}\,e^{u_0 a^\dagger}\,e^{-\bar{u}_0 a},
\end{equation}
\begin{equation}
D(u_0)\equiv D(q_0,p_0)\equiv e^{-\frac{i}{\hbar}q_0 P + \frac{i}{\hbar} p_0 Q}=
e^{\frac{i}{2\hbar}q_0 p_0}\,e^{-\frac{i}{\hbar}q_0 P}\,e^{\frac{i}{\hbar}p_0 Q}=
e^{-\frac{i}{2\hbar}q_0 p_0}\,e^{\frac{i}{\hbar}p_0 Q}\,e^{-\frac{i}{\hbar}q_0 P},
\end{equation}
which is a unitary operator defined over Hilbert space,
\begin{equation}
D^\dagger(u_0)=D^{-1}(u_0)=D(-u_0).
\end{equation}
Indeed the following identities readily follow, which make explicit the displacement action
of the displacement operator $D(u_0)$ on the different quantities being involved,
\begin{equation}
b(u_0)=D(u_0)\,a\,D^\dagger(u_0)= a \,-\, u_0,\quad
b^\dagger(u_0)=D(u_0)\,a^\dagger\,D^\dagger(u_0)=a^\dagger\,-\,\bar{u}_0,
\end{equation}
\begin{equation}
\bar{Q}=D(u_0)\,Q\,D^\dagger(u_0)=Q\,-\,q_0,\qquad
\bar{P}=D(u_0)\,P\,D^\dagger(u_0)=P\,-\,p_0,
\end{equation}
\begin{equation}
D(u_0)\,|q\rangle = e^{\frac{i}{2\hbar}q_0 p_0}\,e^{\frac{i}{\hbar} q p_0}\,|q+q_0\rangle,\qquad
D(u_0)\,|p\rangle = e^{-\frac{i}{2\hbar}q_0 p_0}\,e^{-\frac{i}{\hbar} q_0 p}\,|p+p_0\rangle.
\end{equation}
Consequently the normalised Fock vacuum, $|\Omega_0(u_0)\rangle$, of the displaced Fock algebra, such
that $b(u_0)|\Omega_0(u_0)\rangle=0$ and $\langle\Omega_0(u_0)|\Omega_0(u_0)\rangle=1$,
is obtained as being simply the displaced reference Fock vacuum since $b(u_0)D(u_0)=D(u_0)a$,
\begin{equation}
|\Omega_0(u_0)\rangle=D(u_0)\,|\Omega_0\rangle,\quad
\langle\Omega_0(u_0)|\Omega_0(u_0)\rangle=1,
\end{equation}
\begin{equation}
b(u_0)|\Omega_0(u_0)\rangle=0,\qquad
(a-u_0)|\Omega_0(u_0)\rangle=0,\qquad
a|\Omega_0(u_0)\rangle=u_0|\Omega_0(u_0)\rangle.
\end{equation}
In other words, the Fock vacuum $|\Omega_0(u_0)\rangle$ of the displaced Fock algebra
is a canonical coherent state of the reference Fock vacuum $|\Omega_0\rangle$.
This also implies that all such states $|\Omega_0(u_0)\rangle$ again saturate not only
the SR-UR but also the H-UR with still a vanishing expectation value for the $(Q,P)$
correlator, $\langle\left\{\bar{Q},\bar{P}\right\}\rangle$, but this time with non-vanishing
expectation values for $Q$ and $P$ which are specified by the choice for $u_0$,
\begin{equation}
\langle\Omega_0(u_0)|Q|\Omega_0(u_0)\rangle=q_0,\quad
\langle\Omega_0(u_0)|P|\Omega_0(u_0)\rangle=p_0,\quad
\langle\Omega_0(u_0)|\left\{\bar{Q},\bar{P}\right\}|\Omega_0(u_0)\rangle=0,
\end{equation}
while the values for $\Delta Q$ and $\Delta P$ remain those
of the reference Fock vacuum $|\Omega_0\rangle$,
\begin{equation}
\left(\Delta Q\right)^2=\frac{1}{2}\ell_0^2,\quad
\left(\Delta P\right)^2=\frac{1}{2}\frac{\hbar^2}{\ell_0^2},\quad
\left(\Delta Q\right)^2\,\left(\Delta P\right)^2=\frac{1}{4}\hbar^2.
\end{equation}

As is well known, the coherent states $|\Omega_0(u_0)\rangle$ possess
some remarkable properties\cite{Klauder1,Klauder2}
of which two are worth to be emphasized in our discussion. Even though these states are not
linearly independent among themselves as is made explicit by their non-vanishing overlap matrix elements,
none of which is vanishing,
\begin{eqnarray}
\langle\Omega_0(u_2)|\Omega_0(u_1)\rangle &=& e^{-\frac{1}{2}|u_2|^2-\frac{1}{2}|u_1|^2+\bar{u}_2u_1}
=e^{-\frac{1}{2}(u_2\bar{u}_1-\bar{u}_2 u_1)}\,e^{-\frac{1}{2}|u_2-u_1|^2} \nonumber \\
&=&e^{\frac{i}{2\hbar}(q_2p_1 - q_1 p_2)}\,
e^{-\frac{1}{4\ell_0^2}(q_2-q_1)^2-\frac{1}{4}\left(\frac{\ell_0}{\hbar}\right)^2(p_2-p_1)^2},
\end{eqnarray}
their linear span over all possible values of the parameter $u_0\in\mathbb{C}$ encompasses the
complete Hilbert space of the system. As a matter of fact this latter result remains valid whatever
the choice of normalised reference quantum state on which the displacement operator acts.
Thus given an arbitrary state $|\chi_0\rangle$ normalised to unity, $\langle\chi_0|\chi_0\rangle=1$,
consider the states obtained from the action on it of $D(u_0)$ for all possible values of $u_0\in\mathbb{C}$,
\begin{equation}
|u_0,\chi_0\rangle\equiv D(u_0)\,|\chi_0\rangle.
\end{equation}
One then has the following overcompleteness relation in Hilbert space\footnote{With
$du_0 d\bar{u}_0\equiv d{\rm Re}\,u_0\,d{\rm Im}\,u_0$.},
\begin{equation}
\int_{\mathbb{C}}\frac{du_0\,d\bar{u}_0}{\pi}\,|u_0,\chi_0\rangle\,\langle u_0,\chi_0|=
\int_{\mathbb{R}^2}\frac{dq_0 dp_0}{2\pi\hbar}\,|u_0,\chi_0\rangle\,\langle u_0,\chi_0|=\mathbb{I},
\label{eq:OC1}
\end{equation}
a result which may readily be established by computing the matrix elements of both terms of this
equality in the $Q$ eigenstate basis, for instance. In particular, by choosing for $|\chi_0\rangle$
the reference Fock vacuum $|\Omega_0\rangle$, one obtains the overcompleteness relation
for the displaced Fock vacua $|\Omega_0(u_0)\rangle$,
\begin{equation}
\int_{\mathbb{C}}\frac{du_0 d\bar{u}_0}{\pi}\,|\Omega_0(u_0)\rangle\,\langle\Omega_0(u_0)|=
\int_{\mathbb{R}^2}\frac{dq_0 dp_0}{2\pi\hbar}\,|\Omega_0(u_0)\rangle\,\langle\Omega_0(u_0)|=\mathbb{I}.
\label{eq:OC2}
\end{equation}
This specific result will be shown to extend to all saturating states of the SR-UR.

Another remarkable property of the states $|\Omega_0(u_0)\rangle$ which extends to all saturating states
of the SR-UR is the following. Any finite order polynomial in the Heisenberg observables $Q$ and $P$
possesses a diagonal kernel integral representation in terms of the states $|\Omega_0(u_0)\rangle$, a result which
extends the above overcompleteness relation valid specifically for the unit operator. Let us point out however,
that this property applies specifically for the states $|\Omega_0(u_0)\rangle$ constructed out of the reference
Fock vacuum $|\Omega_0\rangle$. Generically, it does not apply\footnote{Unless of course, one considers
a reference state which itself is again the Fock vacuum for some other Fock algebra constructed out of
the Heisenberg algebra of the observables $Q$ and $P$, as is the case with the squeezed quantum states
to be identified in Section \ref{Sect4}.} for other choices of reference state $|\chi_0\rangle$.

To establish such a result, first consider a general finite order polynomial in the operators $Q$ and $P$.
Such a composite operator may always be brought into the form of a finite sum of normal ordered monomials
relative to the reference Fock algebra $(a,a^\dagger)$. A generating function of such normal ordered
monomials is provided by the operator ${\rm exp}(\alpha a^\dagger)\,{\rm exp}(-\bar{\alpha}a)$
with $\alpha$ and $(-\bar{\alpha}=-\alpha^*)$ as independent generating parameters. Using the above
overcompleteness relation and the fact that $a|\Omega_0(u_0)\rangle=u_0|\Omega_0(u_0)\rangle$,
this generating operator may be given the following integral representation (see also (\ref{eq:BCH2}) in
Appendix B),
\begin{eqnarray}
e^{\alpha a^\dagger}\,e^{-\bar{\alpha}a} &=& e^{|\alpha|^2}\,e^{-\bar{\alpha}a}\,e^{\alpha a^\dagger} \nonumber \\
&=& \int_{\mathbb{C}}\frac{du_0 d\bar{u}_0}{\pi}\,e^{|\alpha|^2}\,e^{-\bar{\alpha}a}\,|\Omega_0(u_0)\rangle\,
\langle\Omega_0(u_0)|\,e^{\alpha a^\dagger}
 \nonumber \\
 &=& \int_{\mathbb{C}}\frac{du_0 d\bar{u}_0}{\pi}\,|\Omega_0(u_0)\rangle\,\left(e^{|\alpha|^2}e^{-\bar{\alpha}u_0} e^{\alpha\bar{u}_0}\right)\,\langle\Omega_0(u_0)|.
\end{eqnarray}
However the product of exponential factors appearing inside this integral is directly related to the diagonal matrix elements of the same operator for the coherent states $|\Omega_0(u_0)\rangle$,
\begin{equation}
\langle\Omega_0(u_0)|e^{\alpha a^\dagger}\,e^{-\bar{\alpha}a}|\Omega_0(u_0)\rangle=
e^{\alpha\bar{u}_0}\,e^{-\bar{\alpha}u_0},\qquad
e^{-\partial_{u_0}\partial_{\bar{u}_0}}\,e^{\alpha\bar{u}_0}\,e^{-\bar{\alpha}u_0}=
e^{\bar{\alpha}\alpha}\,e^{\alpha\bar{u}_0}\,e^{-\bar{\alpha}u_0},
\end{equation}
leading to the final diagonal kernel integral representation of the generating operator,
\begin{equation}
e^{\alpha a^\dagger}\,e^{-\bar{\alpha}a} =
\int_{\mathbb{C}}\frac{du_0 d\bar{u}_0}{\pi}\,|\Omega_0(u_0)\rangle\,
\left(e^{-\partial_{u_0}\partial_{\bar{u}_0}}\,\langle\Omega_0(u_0)|e^{\alpha a^\dagger}\,
e^{-\bar{\alpha}a}|\Omega_0(u_0)\rangle\right)\,
\langle\Omega_0(u_0)|.
\end{equation}
Therefore any composite operator, $A$, which is a finite order polynomial in the observables
$Q$ and $P$ possesses the following diagonal kernel integral representation over the states $|\Omega_0(u_0)\rangle$,
\begin{equation}
A=\int_{\mathbb{C}}\frac{du_0 d\bar{u}_0}{\pi}\,|\Omega_0(u_0)\rangle\,a(u_0,\bar{u}_0)\,\langle\Omega_0(u_0)|,
\end{equation}
where the diagonal kernel $a(u_0,\bar{u}_0)$ is constructed as follows out of the diagonal
matrix elements of $A$ in the states $|\Omega_0(u_0)\rangle$,
\begin{equation}
a(u_0,\bar{u}_0)=e^{-\partial_{u_0}\partial_{\bar{u}_0}}\,A(u_0,\bar{u}_0),\qquad
A(u_0,\bar{u}_0)=\langle\Omega_0(u_0)|A|\Omega_0(u_0)\rangle.
\end{equation}
In terms of the parameters $(q_0,p_0)$, the same results are expressed as, with
$|\Omega_0(q_0,p_0)\rangle\equiv|\Omega_0(u_0)\rangle$,
\begin{equation}
A=\int_{\mathbb{R}^2}\frac{dq_0 dp_0}{2\pi\hbar}\,|\Omega_0(q_0,p_0)\rangle\,a(q_0,p_0)\,\langle\Omega_0(q_0,p_0)|,
\end{equation}
where
\begin{equation}
a(q_0,p_0)={\rm exp}\left(-\frac{1}{2}\ell_0^2\partial^2_{q_0}-\frac{1}{2}\frac{\hbar^2}{\ell^2_0}\partial^2_{p_0}\right)\,
\langle\Omega_0(q_0,p_0)|A|\Omega_0(q_0,p_0)\rangle.
\end{equation}

\section{Fock Algebras for the Saturating Quantum States}
\label{Sect4}

\subsection{Reversible parametrisation packages}
\label{Subsect4.1}

Let us now address the quantum state content, $|\psi_0\rangle$, of the defining relation (\ref{eq:defining1})
for the saturated SR-UR of the Heisenberg algebra of quantum observables $(Q,P)$,
namely,
\begin{equation}
\left[(Q-q_0)\,-\,\lambda_0\,(P-p_0)\right]|\psi_0\rangle=0,
\end{equation}
with
\begin{equation}
\lambda_0=\frac{1}{\left(\Delta P\right)^2}\left(\frac{1}{2}\langle\left\{\bar{Q},\bar{P}\right\}\rangle - \frac{1}{2}i\hbar\right)
=\left(\Delta Q\right)^2\frac{1}{\frac{1}{2}\langle\left\{\bar{Q},\bar{P}\right\}\rangle + \frac{1}{2}i\hbar} .
\end{equation}
Besides the complex variable $u_0$ already representing the real expectation values $(q_0,p_0)$ given
the physical scale $\ell_0$, let us now introduce the angular parameter $\varphi$
related to the possible $(Q,P)$ correlation, such that $-\pi/2\le \varphi\le +\pi/2$ and defined by,
\begin{equation}
\cos\varphi=\frac{1}{\sqrt{1+\frac{1}{\hbar^2}\langle\left\{\bar{Q},\bar{P}\right\}\rangle^2}},\quad
\sin\varphi=\frac{\frac{1}{\hbar}\langle\left\{\bar{Q},\bar{P}\right\}\rangle}
{\sqrt{1+\frac{1}{\hbar^2}\langle\left\{\bar{Q},\bar{P}\right\}\rangle^2}},\quad
\tan\varphi=\frac{1}{\hbar}\langle\left\{\bar{Q},\bar{P}\right\}\rangle.
\end{equation}
Note that the saturated SR-UR is then expressed simply as,
\begin{equation}
\Delta Q\,\Delta P=\frac{\hbar}{2}\,\frac{1}{\cos\varphi}=\frac{\hbar}{2}\sqrt{1+\tan^2\varphi}\ge \frac{\hbar}{2},\quad
\left(\frac{\Delta Q}{\ell_0}\right)\left(\frac{\ell_0}{\hbar}\Delta P\right)=\frac{1}{2\cos\varphi}
=\frac{1}{2}\sqrt{1+\tan^2\varphi}\ge\frac{1}{2},
\end{equation}
while the parameter $\lambda_0$ simplifies to,
\begin{equation}
-\lambda_0=\frac{i\hbar}{2\left(\Delta P\right)^2}\,\frac{1}{\cos\varphi}\,e^{i\varphi}
=i\,\frac{\Delta Q}{\Delta P}\,e^{i\varphi},
\end{equation}
the latter expression thus also displaying explicitly the remaining fourth and last real (and positive) independent
free parameter labelling the saturating states, namely the ratio $\Delta Q/\Delta P$. In particular we then have for the
operator which annihilates the saturating quantum states,
\begin{eqnarray}
&&\frac{1}{\Delta Q}\,\left[(Q-q_0)\,-\,\lambda_0\,(P-p_0)\right] =
\left(\frac{Q-q_0}{\Delta Q}\,+\,ie^{i\varphi}\,\frac{P-p_0}{\Delta P}\right)= \nonumber \\
&=& \frac{1}{\sqrt{2}}\left(\frac{\ell_0}{\Delta Q}+\frac{\hbar\,e^{i\varphi}}{\ell_0 \Delta P}\right)\,b(u_0)\ +\
\frac{1}{\sqrt{2}}\left(\frac{\ell_0}{\Delta Q}-\frac{\hbar\, e^{i\varphi}}{\ell_0 \Delta P}\right)\,b^\dagger(u_0).
\label{eq:annihil1}
\end{eqnarray}

Consequently, when having in mind the reference Fock algebra $(a,a^\dagger)$ and in order to account
for this last variable, $\Delta Q/\Delta P$ or $\Delta P/\Delta Q$,
it proves useful to consider the following further definitions of properly normalised quantities,
with $\rho_\pm\ge 0$ and $-\pi\le \theta_\pm\le +\pi$,
\begin{equation}
\rho_\pm\,e^{i\theta_\pm}\equiv
\frac{1}{2\sqrt{2}\cos\varphi}\left(\frac{\ell_0}{\Delta Q}\,\pm\,\frac{\hbar\,e^{i\varphi}}{\ell_0\Delta P}\right)=
\frac{1}{\sqrt{2}}\,\left(\left(\frac{\ell_0}{\hbar}\Delta P\right)\,\pm\,\left(\frac{\Delta Q}{\ell_0}\right)\,e^{i\varphi}\right),
\end{equation}
so that,
\begin{equation}
\rho_\pm=\frac{1}{\sqrt{2}}\,\sqrt{\left(\frac{\ell_0}{\hbar}\Delta P\right)^2+\left(\frac{\Delta Q}{\ell_0}\right)^2\pm1},
\end{equation}
\begin{equation}
\cos\theta_\pm=\frac{1}{\rho_\pm\sqrt{2}}\,
\left[\left(\frac{\ell_0}{\hbar}\Delta P\right)\,\pm\,\cos\varphi\,\left(\frac{\Delta Q}{\ell_0}\right)\right],\quad
\sin\theta_\pm=\pm\frac{1}{\rho_\pm\sqrt{2}}\,
\sin\varphi\,\left(\frac{\Delta Q}{\ell_0}\right),
\end{equation}
\begin{equation}
\tan\theta_\pm=\frac{\pm\sin\varphi\,\left(\frac{\Delta Q}{\ell_0}\right)}
{\left(\frac{\ell_0}{\hbar}\Delta P\right)\,\pm\,\cos\varphi\,\left(\frac{\Delta Q}{\ell_0}\right)}
=\frac{\sin\varphi}{\cos\varphi\,\pm\,\frac{\ell_0^2}{\hbar}\frac{\Delta P}{\Delta Q}}.
\end{equation}
Since $\rho^2_+ - \rho^2_- =1$, let us introduce finally a real parameter $r$ such that $0\le r< +\infty$, defined by,
\begin{eqnarray}
\cosh r &=& \rho_+=\frac{1}{\sqrt{2}}\,
\sqrt{\left(\frac{\ell_0}{\hbar}\Delta P\right)^2+\left(\frac{\Delta Q}{\ell_0}\right)^2\,+\,1}\,\ge\, 1,\nonumber \\
\sinh r &=& \rho_-=\frac{1}{\sqrt{2}}\,
\sqrt{\left(\frac{\ell_0}{\hbar}\Delta P\right)^2+\left(\frac{\Delta Q}{\ell_0}\right)^2\,-\,1}\,\ge\,0.
\end{eqnarray}
In terms of these quantities, the following notations prove to be useful later on as well,
\begin{equation}
\zeta=e^{i\theta}\,\tanh r,\qquad
z=r\,e^{i\theta},\qquad
e^{i\theta}=-e^{i(\theta_- - \theta_+)}=e^{i(\theta_- - \theta_+ \pm \pi)}.
\end{equation}
Note that the complex variable $z$ takes {\sl a priori\/} all its values in the entire complex plane,
while the complex variable $\zeta$ takes all its values inside the unit disk in the complex plane.

Hence given the physical scale $\ell_0$ and any quantum state saturating the SR-UR, the associated real quantities
$q_0$, $p_0$, $\langle\left\{\bar{Q},\bar{P}\right\}\rangle$, $\Delta Q>0$ and $\Delta P>0$, of which four
are independent because of the property
$(\Delta Q)^2(\Delta P)^2=\hbar^2(1+\langle\left\{\bar{Q},\bar{P}\right\}\rangle^2/\hbar^2)/4$,
determine in a unique manner through the above definitions
the two independent complex quantities $u_0$ and $z$ in the complex plane.
These two complex variables, $u_0$ and $z$, thus label all SR-UR saturating quantum states.

Conversely, given the two complex variables $u_0$ and $z$ taking any values in the complex plane,
in terms of the physical scale $\ell_0$ there corresponds to these a SR-UR saturating quantum state,
$|\psi_0\rangle$, whose relevant expectation values are constructed as follows. On the one hand for the
Heisenberg observables $Q$ and $P$, their expectation values are
\begin{equation}
q_0=\frac{1}{\sqrt{2}}\,\ell_0\left(u_0 + \bar{u}_0\right),\qquad
p_0=-\frac{i\hbar}{\ell_0\sqrt{2}}\,\left(u_0 - \bar{u}_0\right),
\end{equation}
while on the other hand their uncertainties are such that,
\begin{equation}
\left(\frac{\Delta Q}{\ell_0}\right)^2+\left(\frac{\ell_0}{\hbar}\Delta P\right)^2=\cosh 2r \ge 1,\qquad
\left(\frac{\Delta Q}{\ell_0}\right)^2-\left(\frac{\ell_0}{\hbar}\Delta P\right)^2=\cos\theta\,\sinh 2r,
\end{equation}
namely\footnote{Note the identity $\cosh^2 2r - \cos^2\theta\sinh^2 2r=1+\sin^2\theta\sinh^2 2r$.},
\begin{equation}
\left(\frac{\Delta Q}{\ell_0}\right)^2=\frac{1}{2}\left(\cosh 2r\,+\,\cos\theta\,\sinh 2r\right),\quad
\left(\frac{\ell_0}{\hbar}\Delta P\right)^2=\frac{1}{2}\left(\cosh 2r\,-\,\cos\theta\,\sinh 2r\right),
\end{equation}
with thus the saturated SR-UR expressed as
\begin{equation}
\left(\Delta Q\right)^2\left(\Delta P\right)^2=\frac{1}{4}\hbar^2\left(1+\sin^2\theta\,\sinh^2 2r\right).
\end{equation}
Furthermore the $(Q,P)$ correlation of these states $|\psi_0\rangle$ is then determined as,
\begin{equation}
\frac{1}{\hbar}\langle\left\{\bar{Q},\bar{P}\right\}\rangle=\tan\varphi=\sin\theta\,\sinh(2r),
\end{equation}
with
\begin{equation}
\cos\varphi=\frac{1}{\sqrt{1+\sin^2\theta\sinh^2(2r)}},\qquad
\sin\varphi=\frac{\sin\theta\sinh(2r)}{\sqrt{1+\sin^2\theta\sinh^2(2r)}}.
\end{equation}

The particular case of $(Q,P)$ uncorrelated saturating states is worth a separate discussion.
This situation, characterised by the vanishing correlation $\langle\left\{\bar{Q},\bar{P}\right\}\rangle=0$, corresponds
to the phase value $\varphi=0$. One then finds
\begin{equation}
\cos\theta_\pm={\rm sgn}\left(\frac{\ell_0}{\hbar}\Delta P\,\pm\,\frac{\Delta Q}{\ell_0}\right),\qquad
\sin\theta_\pm=0.
\end{equation}
Consequently in such a case $\theta_+=0$, while the value for $\theta_-$ is determined as follows,
\begin{equation}
{\rm if}\quad\frac{\ell_0}{\hbar}\Delta P\,-\,\frac{\Delta Q}{\ell_0}>0:\quad \theta_-=0;\qquad
{\rm if}\quad\frac{\ell_0}{\hbar}\Delta P\,-\,\frac{\Delta Q}{\ell_0}<0:\quad \theta_-=\pm\pi,
\end{equation}
leading to the value for $\theta\equiv\theta_- - \theta_+ \pm \pi$ (mod $2\pi$) given as,
\begin{equation}
{\rm if}\quad\frac{\ell_0}{\hbar}\Delta P\,-\,\frac{\Delta Q}{\ell_0}>0:\quad \theta=\pm\pi\ ({\rm mod}\ 2\pi);\qquad
{\rm if}\quad\frac{\ell_0}{\hbar}\Delta P\,-\,\frac{\Delta Q}{\ell_0}<0:\quad \theta=0\ ({\rm mod}\ 2\pi).
\end{equation}
Nonetheless the value for $r$ remains arbitrary,
\begin{equation}
\cosh r=\frac{1}{\sqrt{2}}\left(\frac{\ell_0}{\hbar}\Delta P\,+\,\frac{\Delta Q}{\ell_0}\right),\quad
\sinh r=\frac{1}{\sqrt{2}}\left|\frac{\ell_0}{\hbar}\Delta P\,-\,\frac{\Delta Q}{\ell_0}\right|.
\end{equation}
On the other hand, in terms of the $(u_0,z)$ parametrisation $(Q,P)$ uncorrelated saturating states
correspond to either one of the two values $\theta=0, \pm\pi$ (mod $2\pi$), thus leading to the quantities,
\begin{eqnarray}
{\rm if}\quad \theta=0: && \frac{\Delta Q}{\ell_0}=\frac{1}{\sqrt{2}}e^r,\quad
\frac{\ell_0}{\hbar}\Delta P=\frac{1}{\sqrt{2}}e^{-r},\quad
\frac{\ell_0}{\hbar}\Delta P - \frac{\Delta Q}{\ell_0}=-\sqrt{2}\sinh r<0; \nonumber \\
{\rm if}\quad \theta=\pm\pi: && \frac{\Delta Q}{\ell_0}=\frac{1}{\sqrt{2}}e^{-r},\quad
\frac{\ell_0}{\hbar}\Delta P=\frac{1}{\sqrt{2}}e^{r},\quad
\frac{\ell_0}{\hbar}\Delta P - \frac{\Delta Q}{\ell_0}=\sqrt{2}\sinh r>0.
\end{eqnarray}
Of course, these results are consistent with those derived above.

Given the latter expressions for $\Delta Q/\ell_0$ and $\ell_0\Delta P/\hbar$, it is clear why the
parameter $r\ge 0$ is known as the squeezing parameter, while all such $(Q,P)$ uncorrelated states
then all saturate the H-UR rather than the SR-UR whatever the value for $r$. In addition, as the value for
the correlation parameter $\theta$ or $\varphi$ varies away from $\theta=0,\pm\pi$ (mod $2\pi$)
or $\varphi=0$ ($-\pi/2\le\varphi\le\pi/2$), respectively, both these quantities remain limited within
a finite interval whose width is set by the squeezing parameter $r$,
\begin{equation}
\frac{1}{\sqrt{2}}\,e^{-r}\,\le\,\frac{\Delta Q}{\ell_0},\,\frac{\ell_0}{\hbar}\Delta P\,\le\,\frac{1}{\sqrt{2}}\,e^{r}.
\end{equation}
In particular when $r=0$, corresponding to $z=0$ and thus to an irrelevant value for $\theta$,
one has the specific situation that $\Delta Q/\ell_0=1/\sqrt{2}=\ell_0\Delta P/\hbar$
in addition to the fact that $\langle\left\{\bar{Q},\bar{P}\right\}\rangle=0$,
namely the fact that $\varphi=0$, thereby leaving as only remaining free parameter the complex variable $u_0$
for these states which saturate the H-UR rather than the SR-UR.
This situation corresponds exactly to the displaced Fock vacua and coherent states
$|\Omega_0(u_0)\rangle=D(u_0)|\Omega_0\rangle$ constructed in Sect.\ref{Sect3} out of the reference
Fock vacuum $|\Omega_0\rangle$.

Consequently in this paper those quantum states that saturate the SR-UR are referred to generally
as ``squeezed states" (or squeezed coherent states, since they turn out to correspond to coherent states
as well, as discussed hereafter). Note that if the parameter $z$ is purely real, whether positive or negative
(thus corresponding to $\theta=0$ or $\theta=\pm\pi$ (mod $2\pi$), respectively),
such squeezed states have no $(Q,P)$ correlation. While if $z$ is
strictly complex with $\theta\ne 0,\pm\pi$ (mod $2\pi$) those squeezed states have a non-vanishing
$(Q,P)$ correlation. If the distinction needs to be emphasized, in this paper these situations will be referred
to as ``uncorrelated" and ``correlated" squeezed states, respectively. Note that uncorrelated squeezed states
saturate the H-UR, while correlated squeezed states saturate the SR-UR but not the H-UR.
In the literature some authors reserve
the term ``squeezed states" specifically to uncorrelated squeezed states thus with $z$ strictly real and
which saturate the H-UR, while to emphasize the distinction correlated squeezed states are then referred
to as ``intelligent states" which saturate the SR-UR\cite{Trifonov,Angelow,Brif}.
However given the considerations of this paper based
on the SR-UR leading to this general class of squeezed states, whether $(Q,P)$ correlated or not
it seems preferable to refer to all of these as squeezed states. Furthermore when time evolution of such states is
considered\footnote{In the simple situation of the harmonic oscillator of mass $m$ and angular frequency $\omega$,
and by choosing then $\ell_0=\sqrt{\hbar/(m\omega)}$, all these general squeezed states evolve coherently into
one another with parameters $u_0$ and $z$ whose time dependence is given by $u_0(t)=u_0e^{i\omega t}$ and
$z(t)=ze^{2i\omega t}$.} the value for $\theta$ certainly evolves in time, thereby generating
correlated squeezed states out of what could have been initially uncorrelated ones.

Another reason why it is legitimate to consider on a same footing correlated and uncorrelated squeezed states
is the following fact. The general Schr\"odinger-Robertson uncertainty relation may also be expressed\cite{Robert}
in terms of the determinant of the covariance matrix of bi-correlations of observables (see Appendix A),
\begin{equation}
\langle\bar{A}^2\rangle\,\langle\bar{B}^2\rangle\,-\,
\langle\bar{A}\bar{B}\rangle\,\langle\bar{B}\bar{A}\rangle\,\ge\,0,\qquad
\left(\Delta A\right)^2\left(\Delta B\right)^2\,-\,\left(\frac{1}{2}\langle\left\{\bar{A},\bar{B}\right\}\rangle\right)^2\,\ge\,
\left(\frac{1}{2}\langle(-i)\left[A,B\right]\rangle\right)^2,
\end{equation}
which in the case of the Heisenberg observables reads,
\begin{equation}
\langle\bar{Q}^2\rangle\,\langle\bar{P}^2\rangle\,-\,
\langle\bar{Q}\bar{P}\rangle\,\langle\bar{P}\bar{Q}\rangle\,\ge\,0,\qquad
\left(\Delta Q\right)^2\left(\Delta P\right)^2\,-\,\hbar^2\left(\frac{1}{2\hbar}
\langle\left\{\bar{Q},\bar{P}\right\}\rangle\right)^2\,\ge\,\frac{1}{4}\hbar^2.
\end{equation}
General squeezed states thus minimize the l.h.s. of these inequalities, whether correlated or uncorrelated,
namely whether the parameter $z$ is strictly complex or strictly real, respectively.

\subsection{Correlated squeezed Fock algebras and their vacua}
\label{Subsect4.2}

Coming back now to the operator (\ref{eq:annihil1}) which annihilates the saturating states, note that it may be
expressed in the form
\begin{equation}
2\cos\varphi\,e^{i\theta_+}\left(\cosh r\,b(u_0)\,-\,e^{i\theta}\,\sinh r\,b^\dagger(u_0)\right)
=2\cos\varphi\,e^{i\theta_+}\,\cosh r\,\left(b(u_0)\,-\,\zeta\,b^\dagger(u_0)\right).
\end{equation}
Consequently let us now introduce the correlated displaced squeezed Fock algebra generators
defined as\footnote{Note the slight abuse of notation which is without consequence, which consists
in denoting as a dependence on $z$ a dependence of $(b(z,u_0),b^\dagger(z,u_0))$ which is in fact separate
in $r$ and in $e^{i\theta}$ while $z=r e^{i\theta}$.}
\begin{equation}
b(z,u_0) = \cosh r\,b(u_0)\,-\,e^{i\theta}\,\sinh r\,b^\dagger(u_0)=
\cosh r\,\left(a-u_0\right)\,-\,e^{i\theta}\,\sinh r\,\left(a^\dagger-\bar{u}_0\right),
\end{equation}
\begin{equation}
b^\dagger(z,u_0) = -e^{-i\theta}\,\sinh r\,b(u_0)\,+\,\cosh r\,b^\dagger(u_0)=
-e^{-i\theta}\,\sinh r\,\left(a-u_0\right)\,+\,\cosh r\,\left(a^\dagger-\bar{u}_0\right),
\end{equation}
which are such that
\begin{equation}
\left[b(z,u_0),b^\dagger(z,u_0)\right]=\mathbb{I},
\end{equation}
while a specific choice of overall phase factor has been effected for $b(z,u_0)$ and $b^\dagger(z,u_0)$,
consistent with the fact that $b(0,u_0)=b(u_0)$ and $b^\dagger(0,u_0)=b^\dagger(u_0)$.

Obviously the SR-UR saturating or squeezed quantum states are the normalised Fock vacua of these displaced
squeezed Fock algebras $(b(z,u_0),b^\dagger(z,u_0))$. Let us denote these Fock states as $|\Omega_z(u_0)\rangle$
such that $\langle\Omega_z(u_0)|\Omega_z(u_0)\rangle=1$ as well as $b(z,u_0)|\Omega_z(u_0)\rangle=0$.
However one also observes that (hence
the name of displaced squeezed Fock algebra for $(b(z,u_0),b^\dagger(z,u_0))$),
\begin{equation}
b(z,u_0)=D(u_0)\,a(z)\,D^\dagger(u_0),\qquad
b^\dagger(z,u_0)= D(u_0)\,a^\dagger(z)\,D^\dagger(u_0),
\end{equation}
where the operators
\begin{equation}
a(z)= \cosh r\,a\,-\,e^{i\theta}\,\sinh r\,a^\dagger,\qquad
a^\dagger(z)= -e^{-i\theta}\,\sinh r\,a\,+\,\cosh r\,a^\dagger,
\label{eq:Bogo1}
\end{equation}
define correlated squeezed Fock algebras such that
\begin{equation}
\left[a(z),a^\dagger(z)\right]=\mathbb{I},
\end{equation}
which are general Bogoliubov transformations of the reference Fock algebra $(a,a^\dagger)$ such that
$(a(0),a^\dagger(0))=(a,a^\dagger)$.
Consequently if $|\Omega_z\rangle$ denote the normalised Fock vacua of the Fock algebras $(a(z),a^\dagger(z))$
for all $z\in\mathbb{C}$, such that $\langle\Omega_z|\Omega_z\rangle=1$ and $a(z)|\Omega_z\rangle=0$,
the saturating squeezed states and thus also Fock vacua $|\Omega_z(u_0)\rangle$
are given as the displaced states of $|\Omega_z\rangle$,
\begin{equation}
|\Omega_z(u_0)\rangle = D(u_0)\,|\Omega_z\rangle,\qquad
b(z,u_0)\,|\Omega_z(u_0)\rangle=D(u_0)\,a(z)\,|\Omega_z\rangle=0.
\end{equation}
Note that from the last of these two identities it follows that general squeezed states $|\Omega_z(u_0)\rangle$
with $u_0\ne 0$ are also coherent states of the squeezed $(a(z),a^\dagger(z))$ Fock algebras.
Indeed by introducing the quantities
\begin{eqnarray}
u_0(z) &\equiv& \cosh r\,u_0\,-\,e^{i\theta}\,\sinh r\,\bar{u}_0 = \cosh r\,\left(u_0\,-\,\zeta\,\bar{u}_0\right),\quad
u_0(0)=u_0,\nonumber \\
\bar{u}_0(z) &\equiv& -e^{-i\theta}\,\sinh r\,u_0\,+\,\cosh r\,\bar{u}_0=
\cosh r\,\left(\bar{u}_0\,-\,\bar{\zeta}\,u_0\right),\quad
\bar{u}_0(0)=\bar{u}_0,
\end{eqnarray}
one has,
\begin{equation}
a(z)\,|\Omega_z(u_0)\rangle=u_0(z)\,|\Omega_z(u_0)\rangle,
\end{equation}
as follows also from the identity, 
\begin{equation}
b(z,u_0)=a(z)\,-\,u_0(z),
\end{equation}
which shows that the $(b(z,u_0),b^\dagger(z,u_0))$ Fock algebras are
shifted versions of the $(a(z),a^\dagger(z))$ Fock algebras\footnote{In the same way that
$b(u_0)|\Omega_0(u_0)\rangle=0$, $a|\Omega_0(u_0)\rangle=u_0|\Omega_0(u_0)\rangle$ and $b(u_0)=a-u_0$,
corresponding to the case with $z=0$.}.
As a matter of fact it may readily be checked that one has, independently from the value for $z$,
\begin{equation}
u_0(z)a^\dagger(z)\,-\,\bar{u}_0(z) a(z)=u_0 a^\dagger\,-\,\bar{u}_0 a,
\label{eq:ident2}
\end{equation}
so that,
\begin{equation}
D(u_0)=e^{-\frac{i}{\hbar}q_0 P + \frac{i}{\hbar} p_0 Q}=e^{u_0a^\dagger - \bar{u}_0 a}=
e^{u_0(z)a^\dagger(z)\,-\,\bar{u}_0(z) a(z)},
\end{equation}
a property which thus explains the above results.

Inverting the Bogoliubov transformations (\ref{eq:Bogo1}), one finds,
\begin{equation}
a=\cosh r\,a(z)\,+\,e^{i\theta}\,\sinh r\,a^\dagger(z),\qquad
a^\dagger=e^{-i\theta}\,\sinh r\,a(z)\,+\,\cosh r\,a^\dagger(z),
\end{equation}
hence likewise for the variables $u_0$ and $u_0(z)$,
\begin{equation}
u_0=\cosh r\,u_0(z)\,+\,e^{i\theta}\,\sinh r\,\bar{u}_0(z),\qquad
\bar{u}_0=e^{-i\theta}\,\sinh r\,u_0(z)\,+\,\cosh r\,\bar{u}_0(z).
\end{equation}
In terms of the Heisenberg observables, these definitions translate into,
\begin{eqnarray}
a(z) &=& \frac{1}{\sqrt{2}}\left(\cosh r - e^{i\theta}\,\sinh r\right)\frac{Q}{\ell_0}
+\frac{i}{\sqrt{2}}\left(\cosh r + e^{i\theta}\,\sinh r\right)\frac{\ell_0}{\hbar}P, \nonumber \\
a^\dagger(z) &=& \frac{1}{\sqrt{2}}\left(\cosh r - e^{-i\theta}\,\sinh r\right)\frac{Q}{\ell_0}
-\frac{i}{\sqrt{2}}\left(\cosh r + e^{-i\theta}\,\sinh r\right)\frac{\ell_0}{\hbar}P,
\end{eqnarray}
with the inverse relations,
\begin{eqnarray}
\frac{Q}{\ell_0} &=& \frac{1}{\sqrt{2}}\left(\cosh r + e^{-i\theta}\,\sinh r\right)\,a(z)\,+\,
\frac{1}{\sqrt{2}}\left(\cosh r + e^{i\theta}\,\sinh r\right)\,a^\dagger(z), \nonumber \\
\frac{\ell_0}{\hbar}P &=& -\frac{i}{\sqrt{2}}\left(\cosh r - e^{-i\theta}\,\sinh r\right)\,a(z)\,+\,
\frac{i}{\sqrt{2}}\left(\cosh r - e^{i\theta}\,\sinh r\right)\,a^\dagger(z),
\end{eqnarray}
so that for the corresponding parameters $u_0(z)$, $\bar{u}_0(z)$, $q_0$ and $p_0$,
\begin{eqnarray}
u_0(z) &=& \frac{1}{\sqrt{2}}\left(\cosh r - e^{i\theta}\,\sinh r\right)\frac{q_0}{\ell_0}
+\frac{i}{\sqrt{2}}\left(\cosh r + e^{i\theta}\,\sinh r\right)\frac{\ell_0}{\hbar}p_0, \nonumber \\
\bar{u}_0(z) &=& \frac{1}{\sqrt{2}}\left(\cosh r - e^{-i\theta}\,\sinh r\right)\frac{q_0}{\ell_0}
-\frac{i}{\sqrt{2}}\left(\cosh r + e^{-i\theta}\,\sinh r\right)\frac{\ell_0}{\hbar}p_0,
\end{eqnarray}
while
\begin{eqnarray}
\frac{q_0}{\ell_0} &=& \frac{1}{\sqrt{2}}\left(\cosh r + e^{-i\theta}\,\sinh r\right)\,u_0(z)\,+\,
\frac{1}{\sqrt{2}}\left(\cosh r + e^{i\theta}\,\sinh r\right)\,\bar{u}_0(z), \nonumber \\
\frac{\ell_0}{\hbar}p_0 &=& -\frac{i}{\sqrt{2}}\left(\cosh r - e^{-i\theta}\,\sinh r\right)\,u_0(z)\,+\,
\frac{i}{\sqrt{2}}\left(\cosh r - e^{i\theta}\,\sinh r\right)\,\bar{u}_0(z).
\end{eqnarray}

\subsection{Squeezed Fock vacua and SR-UR saturating quantum states}
\label{Subsect4.3}

Having understood that the SR-UR saturating states are the displaced coherent states of the squeezed
Fock vacua $|\Omega_z\rangle$, namely $|\Omega_z(u_0)\rangle=D(u_0)|\Omega_z\rangle$,
let us finally turn to the construction of the latter which are
characterised by the condition that $a(z)|\Omega_z\rangle=0$ with
\begin{equation}
a(z)= \cosh r\,a\,-\,e^{i\theta}\,\sinh r\,a^\dagger.
\end{equation}
Given that the corresponding Bogoliubov transformation, linear in both generators of the
reference Fock algebra $(a,a^\dagger)$, is unitary, necessarily it corresponds to a unitary
operator acting on Hilbert space of the following form, defined up to an arbitrary global phase factor
set here to a trivial value,
\begin{equation}
S(\alpha)={\rm exp}\left(\frac{1}{2}\alpha{a^\dagger}^2\,-\,\frac{1}{2}\bar{\alpha}a^2\right),\qquad \alpha\in\mathbb{C},
\end{equation}
$\alpha$ being some complex parameter. The operator $S(\alpha)$ is thus such that
\begin{equation}
S^\dagger(\alpha)=S^{-1}(\alpha)=S(-\alpha),\qquad S(0)=\mathbb{I}.
\end{equation}
A straightforward application of the BCH formula (\ref{eq:BCH1}) in Appendix B then leads to the identities,
\begin{equation}
S(\alpha)\,a\,S^\dagger(\alpha)=\cosh\rho\,a\,-\,e^{i\phi}\,\sinh\rho\,a^\dagger,\quad
S(\alpha)\,a^\dagger\,S^\dagger(\alpha)=-\,e^{-i\phi}\,\sinh\rho\,a\,+\,\cosh\rho\,a^\dagger,
\end{equation}
the parameter $\alpha$ being represented as $\alpha=\rho\,e^{i\phi}$ with $\rho\ge 0$.

Consequently by choosing $\alpha=z$, one finds for the squeezed Fock algebras $(a(z),a^\dagger(z))$,
\begin{equation}
a(z)=S(z)\,a\,S^\dagger(z),\qquad
a^\dagger(z)=S(z)\,a^\dagger\,S^\dagger(z),\qquad
\left[a(z),a^\dagger(z)\right]=\mathbb{I},
\end{equation}
while their normalised squeezed Fock vacua $|\Omega_z\rangle$ are constructed as follows
out of the reference Fock vacuum $|\Omega_0\rangle$, since $a(z)S(z)|\Omega_0\rangle=S(z)a|\Omega_0\rangle=0$,
\begin{equation}
|\Omega_z\rangle = S(z)\,|\Omega_0\rangle,\qquad
\langle\Omega_z|\Omega_z\rangle=\langle\Omega_0|\Omega_0\rangle=1.
\end{equation}
Given the displacement operator $D(u_0)$, let us also introduce the operators
\begin{equation}
S(z,u_0)\equiv {\rm exp}\left(\frac{1}{2}z(a^\dagger-\bar{u}_0)^2-\frac{1}{2}\bar{z}(a-u_0)^2\right),
\end{equation}
which obey the following properties\footnote{Note that because of (\ref{eq:ident2}), one also has
the identity $D(u_0)S(z)=S(z)D(u_0(z))$ with $u_0(z)=\cosh r(u_0-\zeta\bar{u}_0)$. The author
thanks Victor Massart for a remark on this point.},
\begin{equation}
D(u_0)\,S(z)=S(z,u_0)\,D(u_0),\qquad
S(z,u_0)=D(u_0)\,S(z)\,D^\dagger(u_0).
\end{equation}

Hence finally all normalised quantum states that saturate the Schr\"odinger-Robertson uncertainty relation
for the Heisenberg observables $(Q,P)$ are given by the algebraic representation
\begin{eqnarray}
|\psi_0(z,u_0)\rangle &\equiv& |\Omega_z(u_0)\rangle
=e^{u_0a^\dagger - \bar{u}_0 a}\,e^{\frac{1}{2}z{a^\dagger}^2-\frac{1}{2}\bar{z}a^2}\,|\Omega_0\rangle \nonumber \\
&=& D(u_0)\,S(z)\,|\Omega_0\rangle= S(z,u_0)\,D(u_0)\,|\Omega_0\rangle
=S(z)D(u_0(z))\,|\Omega_0\rangle,
\end{eqnarray}
$|\Omega_0\rangle$ being the normalised Fock vacuum of the reference Fock algebra $(a,a^\dagger)$.
Note that this construction also fixes the absolute phase factor for all these saturating states,
relative to the choice of phase made for the state $|\Omega_0\rangle$. The overall phase factor
for the wave function of the saturating states, $\psi_0(q;z,u_0)\equiv\langle q|\psi_0(z,u_0)\rangle$ in
Eq.(\ref{eq:wf1}), will be determined accordingly in Sect.\ref{Sect6}.

\section{Overcompleteness and Kernel Representation}
\label{Sect5}

In Sect.\ref{Sect3} two remarkable properties of the canonical coherent states, $|\Omega_0(u_0)\rangle$,
were emphasized. Let us now consider how these properties extend to the general squeezed
coherent states $|\Omega_z(u_0)\rangle$, beginning with the overcompleteness property.

As established in Eq.(\ref{eq:OC1}), given any normalised reference state $|\chi_0\rangle$,
one has the following representation of the unit operator on the considered Hilbert space,
\begin{equation}
\int_{\mathbb{C}}\frac{du_0\,d\bar{u}_0}{\pi}\,D(u_0)|\chi_0\rangle\,\langle \chi_0|D^\dagger(u_0)=
\int_{\mathbb{R}^2}\frac{dq_0 dp_0}{2\pi\hbar}\,D(q_0,p_0)|\chi_0\rangle\,\langle\chi_0|D^\dagger(q_0,p_0)
=\mathbb{I}.
\end{equation}
Hence by choosing $|\chi_0\rangle=|\Omega_z\rangle$ so that $D(u_0)|\chi_0\rangle=|\Omega_z(u_0)\rangle$,
given any fixed value for $z\in\mathbb{C}$ one has the overcompleteness property for the SR-UR saturating states,
\begin{equation}
\int_{\mathbb{C}}\frac{du_0\,d\bar{u}_0}{\pi}\,|\Omega_z(u_0)\rangle\,\langle \Omega_z(u_0)|=
\int_{\mathbb{R}^2}\frac{dq_0 dp_0}{2\pi\hbar}\,|\Omega_z(u_0)\rangle\,\langle\Omega_z(u_0)|=\mathbb{I},
\label{eq:OC3}
\end{equation}
which thus generalises the overcompleteness relation in Eq.(\ref{eq:OC2}) (which corresponds to the case $z=0$).
Note well however, that this identity involves an integral over the entire complex plane only for the
complex variable $u_0$, independently of the value for $z$ which is fixed but arbitrary. Since the
states $|\Omega_z\rangle=S(z)|\Omega_0\rangle$ involve those Fock states built from the reference
Fock algebra $(a,a^\dagger)$ which include only an even number of the corresponding $a^\dagger$ Fock quanta
and thereby span only half the Hilbert space under consideration, a similar identity involving rather such an integral
only over the complex plane of $z$ values but with a fixed value now for $u_0$, cannot apply.
However, given any arbitrary normalisable and normalised integration measure $\mu(z,\bar{z})$
over the complex plane for all values of $z$, provides still for a generalised form of overcompleteness
relation involving then all the saturating states,
\begin{equation}
\int_{\mathbb{C}^2}\frac{du_0\,d\bar{u}_0}{\pi}\,\frac{dz\,d\bar{z}}{\pi}\,\mu(z,\bar{z})\,
|\Omega_z(u_0)\rangle\,\langle \Omega_z(u_0)|=\mathbb{I},\qquad
\int_{\mathbb{C}}\frac{dz\,d\bar{z}}{\pi}\,\mu(z,\bar{z})=1.
\label{eq:OC4}
\end{equation}

Let us now consider the possibility of a diagonal kernel integral representation of operators.
Given a fixed but arbitrary value for $z$, any finite order polynomial in the observables $Q$ and $P$
may be written as a linear combination of monomials which are expressed in normal ordered
form with respect to the Fock algebra $(a(z),a^\dagger(z))$. Let us thus consider again
the generating function of such normal ordered monomials, namely the
operator ${\rm exp}(\alpha a^\dagger(z)){\rm exp}(-\bar{\alpha}a(z))$ with generating parameters
$\alpha\in\mathbb{C}$ and $(-\bar{\alpha}=-\alpha^*)$. Following the same line of analysis as in
Sect.\ref{Sect3}, one has,
\begin{eqnarray}
e^{\alpha a^\dagger(z)}\,e^{-\bar{\alpha}a(z)} &=&
e^{\alpha\bar{\alpha}}\,e^{-\bar{\alpha}a(z)}\,e^{\alpha a^\dagger(z)} \nonumber \\
&=& \int_{\mathbb{C}}\frac{du_0\,d\bar{u}_0}{\pi}\,e^{\alpha\bar{\alpha}}\,
e^{-\bar{\alpha}a(z)}\,|\Omega_z(u_0)\rangle\,\langle\Omega_z(u_0)|\,e^{\alpha a^\dagger(z)} \nonumber \\
&=& \int_{\mathbb{C}}\frac{du_0\,d\bar{u}_0}{\pi}\,|\Omega_z(u_0)\rangle\,
e^{\alpha\bar{\alpha}}\,e^{-\bar{\alpha}u_0(z)}\,e^{\alpha\bar{u}_0(z)}\,\langle\Omega_z(u_0)|  \\
&=& \int_{\mathbb{C}}\frac{du_0\,d\bar{u}_0}{\pi}\,|\Omega_z(u_0)\rangle\,
\left[e^{-\partial_{u_0(z)}\partial_{\bar{u}_0(z)}}\,\langle\Omega_z(u_0)|
e^{\alpha a^\dagger(z)}\,e^{-\bar{\alpha} a(z)}\,|\Omega_z(u_0)\rangle\right]\,\langle\Omega_z(u_0)|. \nonumber
\end{eqnarray}
Consequently, any finite order polynomial in the Heisenberg observables $Q$ and $P$ may be given
the following diagonal kernel integral representation, whatever the fixed but arbitrary value for the complex
squeezing parameter $z$,
\begin{equation}
A=\int_{\mathbb{C}}\frac{du_0\,d\bar{u}_0}{\pi}\,|\Omega_z(u_0)\rangle\,a(z,\bar{z};u_0,\bar{u}_0)\,
\langle\Omega_z(u_0)|,
\end{equation}
where the diagonal kernel is defined as,
\begin{equation}
a(z,\bar{z};u_0,\bar{u}_0)=e^{-\partial_{u_0(z)}\partial_{\bar{u}_0(z)}}\,
\langle\Omega_z(u_0)|A|\Omega_z(u_0)\rangle.
\end{equation}
More generally given the normalised integration measure $\mu(z,\bar{z})$, one may extend this representation to,
\begin{equation}
A=\int_{\mathbb{C}^2}\frac{du_0\,d\bar{u}_0}{\pi}\,\frac{dz\,d\bar{z}}{\pi}\,\mu(z,\bar{z})\,
|\Omega_z(u_0)\rangle\,a(z,\bar{z};u_0,\bar{u}_0)\,
\langle\Omega_z(u_0)|.
\end{equation}
In the above representations the second order differential operator $\partial_{u_0(z)}\partial_{\bar{u}_0(z)}$
may also be expressed as,
\begin{eqnarray}
&&\partial_{u_0(z)}\partial_{\bar{u}_0(z)} =
\frac{1}{2}e^{i\theta}\sinh 2r\,\partial^2_{u_0}\,+\,\frac{1}{2}e^{-i\theta}\sinh 2r\,\partial^2_{\bar{u}_0}\,+\,
\cosh 2r\,\partial_{u_0}\partial_{\bar{u}_0}  \\
&=& (\cosh 2r + \cos\theta \sinh 2r)\frac{1}{2}\ell^2_0\,\partial^2_{q_0}\,+\,
(\cosh 2r \,-\, \cos\theta\sinh 2r)\frac{1}{2}\frac{\hbar^2}{\ell^2_0}\,\partial^2_{p_0}\,+\,
\hbar\sin\theta\,\partial_{q_0}\,\partial_{p_0}, \nonumber
\end{eqnarray}
while it is worth noting that
\begin{equation}
du_0\,d\bar{u}_0=du_0(z)\,d\bar{u}_0(z).
\end{equation}
Indeed, since (see (\ref{eq:ident2}))
\begin{equation}
|\Omega_z(u_0)\rangle=D(u_0)|\Omega_z\rangle=e^{u_0(z)a^\dagger(z)-\bar{u}_0(z)a(z)}\,|\Omega_z\rangle
=e^{-\frac{1}{2}|u_0(z)|^2}\,e^{u_0(z)a^\dagger(z)}\,|\Omega_z\rangle,
\end{equation}
the matrix element $\langle\Omega_z(u_0)|A|\Omega_z(u_0)\rangle$ is first a function
of $u_0(z)$ and $\bar{u}_0(z)$ rather than directly a function of $u_0$ and $\bar{u}_0$
independently of the value for $z$.

\section{Correlated Squeezed State Wavefunctions}
\label{Sect6}

\subsection{Squeezed state configuration space wave functions}

Having fully identified, in the form recalled hereafter, the quantum states that saturate the Schr\"odinger-Robertson
uncertainty relation for the Heisenberg algebra of the observables $Q$ and $P$,
inclusive of their phase since that of the reference Fock vacuum has been specified,
\begin{equation}
|\psi_0(z,u_0)\rangle=|\Omega_z(u_0)\rangle=D(u_0)\,S(z)\,|\Omega_0\rangle,
\end{equation}
we may reconsider the construction of the wave function representation of these states, say in configuration space.

In terms now of the notations and parametrisations introduced throughout the discussion,
the expression for the wave functions of these states as determined in (\ref{eq:wf1}) and (\ref{eq:N0})
reads\footnote{Since one has the relations
$\frac{i}{2\lambda_0\hbar}=-\frac{1}{\hbar}\frac{\Delta P}{\Delta Q}e^{-i\varphi}=
-\frac{1}{2\ell^2_0}\frac{1-i\sin\theta\sinh 2r}{\cosh 2r + \cos\theta\sinh 2r}$.},
\begin{eqnarray}
\psi_0(q;z,u_0) &\equiv& \langle q|\Omega_z(u_0)\rangle=
\left(\pi\ell^2_0\right)^{-1/4}\,\left(\cosh 2r + \cos\theta\sinh 2r\right)^{-1/4}\,\times \nonumber \\
&&\times\,e^{i\varphi(z,u_0)}\,e^{\frac{i}{\hbar}qp_0}\,
{\rm exp}\left(-\frac{1}{2}\frac{1-i\sin\theta\sinh 2r}{\cosh 2r + \cos\theta\sinh 2r}\left(\frac{q-q_0}{\ell_0}\right)^2\right),
\label{eq:wf2}
\end{eqnarray}
where $\varphi(z,u_0)$ is the phase factor still to be determined. Thus in particular, when $u_0=0$,
\begin{eqnarray}
\langle q|\Omega_z(0)\rangle &=& \langle q|\Omega_z\rangle =
\left(\pi\ell^2_0\right)^{-1/4}\,\left(\cosh 2r + \cos\theta\sinh 2r\right)^{-1/4}\,\times \nonumber \\
&&\times\,e^{i\varphi(z,0)}\,
{\rm exp}\left(-\frac{1}{2}\frac{1-i\sin\theta\sinh 2r}{\cosh 2r + \cos\theta\sinh 2r}\left(\frac{q}{\ell_0}\right)^2\right).
\label{eq:wf3}
\end{eqnarray}
However since the displacement operator's action on $Q$ eigenstates is such that
\begin{equation}
D(u_0)|q\rangle=e^{\frac{i}{2\hbar}q_0p_0}\,e^{\frac{i}{\hbar}qp_0}\,|q+q_0\rangle,\quad
\langle q|D(u_0)=\langle q|D^\dagger(-u_0)=\langle q-q_0|\,e^{-\frac{i}{2\hbar}q_0p_0}\,e^{\frac{i}{\hbar}qp_0},
\end{equation}
one has,
\begin{equation}
\langle q|\Omega_z(u_0)\rangle=\langle q|D(u_0) S(z)|\Omega_0\rangle
=e^{-\frac{i}{2\hbar}q_0p_0}\,e^{\frac{i}{\hbar}q p_0}\,\langle q-q_0|\Omega_z(0)\rangle,
\end{equation}
which, given the above two expressions for $\langle q|\Omega_z(u_0)\rangle$ and $\langle q|\Omega_z(0)\rangle$,
thus implies that
\begin{equation}
e^{i\varphi(z,u_0)}=e^{-\frac{i}{2\hbar}q_0 p_0}\,e^{i\varphi(z,0)}.
\end{equation}

The final determination of the phase factor $\varphi(z,0)$ is based now on the following relation between
specific Fock state overlaps,
\begin{equation}
\langle\Omega_0|\Omega_z\rangle=\int^{+\infty}_{-\infty} dq\,\langle\Omega_0|q\rangle\,\langle q|\Omega_z\rangle.
\label{eq:over1}
\end{equation}
The function $\langle q|\Omega_z\rangle$ is specified in (\ref{eq:wf3}) in terms of $e^{i\varphi(z,0)}$, while given
the choice of phase for the reference Fock vacuum $|\Omega_0\rangle$ its own wave function was
determined earlier on to be simply,
\begin{equation}
\langle q|\Omega_0\rangle=\left(\pi\ell_0^2\right)^{-1/4}\,e^{-\frac{1}{2}\frac{q^2}{\ell^2_0}}.
\end{equation}
On the other hand since the l.h.s. of the overlap (\ref{eq:over1}) corresponds
to $\langle\Omega_0|S(z)|\Omega_0\rangle$, clearly this latter quantity does not involve any phase
factor left unspecified. Consequently the Gaussian integration in (\ref{eq:over1}) determines
the overall phase factor $\varphi(z,u_0)$ of the wave functions (\ref{eq:wf2}).

The evaluation of $\langle\Omega_0|\Omega_z\rangle$ is readily achieved by using the BCH formula
(\ref{eq:BCH-Sz}) of Appendix~B for the squeezing operator $S(z)$,
\begin{equation}
S(z)=e^{\frac{1}{2}\zeta\,{a^\dagger}^2}\,e^{\ln(1-|\zeta|^2)\frac{1}{2}(a^\dagger a+\frac{1}{2})}\,
e^{-\frac{1}{2}\bar{\zeta}a^2},\qquad \zeta=e^{i\theta}\,\tanh r,\quad z=r e^{i\theta}.
\label{eq:Sz}
\end{equation}
Hence,
\begin{equation}
\langle\Omega_0|\Omega_z\rangle=\langle\Omega_0|S(z)|\Omega_0\rangle=
\left(1-\tanh^2 r\right)^{1/4}=\left(\cosh r\right)^{-1/2}.
\end{equation}

When combined with the normalisation factor in (\ref{eq:wf3}), the Gaussian integration in (\ref{eq:over1})
leads to a factor which may be brought into the form of this last factor $(\cosh r)^{-1/2}$ being multiplied
by a specific phase factor. In order to express the thereby determined phase factor $\varphi(z,0)$,
let us introduce two last angular parameters $\bar{\theta}_\pm(z)$ defined by,
\begin{equation}
\cos\bar{\theta}_\pm(z)=\frac{\cosh r \pm \cos\theta \sinh r}{\sqrt{\cosh 2r \pm \cos\theta \sinh 2r}},\qquad
\sin\bar{\theta}_\pm(z)=\frac{\pm\sin\theta \sinh r}{\sqrt{\cosh 2r \pm \cos\theta \sinh 2r}},
\end{equation}
\begin{equation}
\tan\bar{\theta}_\pm(z)=\frac{\pm\sin\theta \sinh r}{\cosh r \pm \cos\theta \sinh r},
\end{equation}
and such that,
\begin{eqnarray}
\cos(\bar{\theta}_+(z)-\bar{\theta}_-(z)) &=& \frac{1}{\sqrt{\cosh^2 2r -\cos^2\theta\sinh^2 2r}},
\nonumber \\
\sin(\bar{\theta}_+(z)-\bar{\theta}_-(z)) &=& \frac{\sin\theta\sinh 2r}{\sqrt{\cosh^2 2r -\cos^2\theta\sinh^2 2r}}.
\end{eqnarray}
On completing the Gaussian integration in (\ref{eq:over1}) (which requires some little work for simplifying
some intermediate expressions), one then finally determines that
\begin{equation}
e^{i\varphi(z,0)} = e^{-\frac{i}{2}\bar{\theta}_+(z)}.
\end{equation}

In conclusion the complete expression for the configuration space wave function representations
of all the states that saturate the Schr\"odinger-Robertson uncertainty relation for the Heisenberg
observables $Q$ and $P$ is given as,
\begin{eqnarray}
\psi_0(q;z,u_0) &\equiv& \langle q|\Omega_z(u_0)\rangle=
\left(\pi\ell^2_0\right)^{-1/4}\,\left(\cosh 2r + \cos\theta\sinh 2r\right)^{-1/4}\,e^{-\frac{i}{2}\bar{\theta}_+(z)}\,\times \nonumber \\
&&\times\,e^{-\frac{i}{2\hbar}q_0 p_0}\,e^{\frac{i}{\hbar}qp_0}\,
{\rm exp}\left(-\frac{1}{2}\frac{1-i\sin\theta\sinh 2r}{\cosh 2r + \cos\theta\sinh 2r}\left(\frac{q-q_0}{\ell_0}\right)^2\right).
\qquad
\label{eq:wf4}
\end{eqnarray}
Note that because of the following identities,
\begin{eqnarray}
\cosh r \pm e^{i\theta}\sinh r=\sqrt{\cosh 2r \pm \cos\theta\sinh 2r}\,e^{i\bar{\theta}_\pm}, \nonumber \\
\cosh r \pm e^{-i\theta}\sinh r=\sqrt{\cosh 2r \pm \cos\theta\sinh 2r}\,e^{-i\bar{\theta}_\pm},
\end{eqnarray}
\begin{eqnarray}
1 \pm i\sin\theta\sinh 2r &=& \left(\cosh r \pm e^{i\theta}\sinh r\right)\,\left(\cosh r \mp e^{-i\theta}\sinh r\right),
\nonumber \\
\cosh 2r \pm \cos\theta\sinh 2r &=& \left(\cosh r \pm e^{i\theta} \sinh r\right)\,\left(\cosh r \pm e^{-i\theta}\sinh r\right),
\end{eqnarray}
the same wave functions have also the equivalent representations,
\begin{eqnarray}
\psi_0(q;z,u_0) &\equiv& \langle q|\Omega_z(u_0)\rangle=
\left(\pi\ell^2_0\right)^{-1/4}\,\left(\cosh 2r + \cos\theta\sinh 2r\right)^{-1/4}\,e^{-\frac{i}{2}\bar{\theta}_+(z)}\,\times \nonumber \\
&&\times\,e^{-\frac{i}{2\hbar}q_0 p_0}\,e^{\frac{i}{\hbar}qp_0}\,
{\rm exp}\left(-\frac{1}{2}\frac{\cosh r - e^{i\theta}\sinh r}{\cosh r + e^{i\theta}\sinh r}\left(\frac{q-q_0}{\ell_0}\right)^2\right),
\label{eq:wf5}
\end{eqnarray}
and
\begin{eqnarray}
\psi_0(q;z,u_0) &\equiv& \langle q|\Omega_z(u_0)\rangle=
\left(\pi\ell^2_0\right)^{-1/4}\,\left(\cosh 2r + \cos\theta\sinh 2r\right)^{-1/4}\,e^{-\frac{i}{2}\bar{\theta}_+(z)}\,\times  \\
&&\times\,e^{-\frac{i}{2\hbar}q_0 p_0}\,e^{\frac{i}{\hbar}qp_0}\,
{\rm exp}\left(-\frac{1}{2}\sqrt{\frac{\cosh 2r - \cos\theta\sinh 2r}{\cosh 2r + \cos\theta\sinh 2r}}\,
e^{i(\bar{\theta}_-(z)-\bar{\theta}_+(z))}\left(\frac{q-q_0}{\ell_0}\right)^2\right). \nonumber
\label{eq:wf6}
\end{eqnarray}
Furthermore note that,
\begin{equation}
\left(\cosh 2r \pm\cos\theta\sinh 2r\right)^{-1/4}\,e^{-\frac{i}{2}\bar{\theta}_\pm(z)}=
\left(\cosh r \pm e^{i\theta}\sinh r\right)^{-1/2},
\end{equation}
a relation which invites us to consider finally the result in the following form,
\begin{eqnarray}
\psi_0(q;z,u_0) &\equiv& \langle q|\Omega_z(u_0)\rangle=
\left(\pi\ell^2_0\right)^{-1/4}\,\left(\cosh r + e^{i\theta}\sinh r\right)^{-1/2}\,\times  \\
&&\times\,e^{-\frac{i}{2\hbar}q_0 p_0}\,e^{\frac{i}{\hbar}qp_0}\,
{\rm exp}\left(-\frac{1}{2}\frac{\cosh r - e^{i\theta}\sinh r}{\cosh r + e^{i\theta}\sinh r}\left(\frac{q-q_0}{\ell_0}\right)^2\right). \nonumber
\label{eq:wf7}
\end{eqnarray}

\subsection{The fundamental overlap $\langle\Omega_{z_2}(u_2)|\Omega_{z_1}(u_1)\rangle$ of squeezed states}

As a last quantity to be determined in this paper, let us consider the overlap of two arbitrary general
squeezed coherent states, associated to the pairs of variables $(z_2,u_2)$ and $(z_1,u_1)$,
\begin{equation}
\langle\Omega_{z_2}(u_2)|\Omega_{z_1}(u_1)\rangle=\int_{-\infty}^{+\infty}dq\,\langle\Omega_{z_2}(u_2)|q\rangle\,
\langle q|\Omega_{z_1}(u_1)\rangle.
\label{eq:over2}
\end{equation}
Given the parameters $u_2$ and $u_1$, correspondingly one has the pairs of quantities $(q_2,p_2)$
and $(q_1,p_1)$, while related to $z_2$ and $z_1$ one has the remaining variables $(r_2,\theta_2)$
and $(r_1,\theta_1)$ such that
\begin{equation}
z_2=r_2\,e^{i\theta_2},\quad
\zeta_2=e^{i\theta_2}\,\tanh r_2;\quad
z_1=r_1\,e^{i\theta_1},\quad
\zeta_1=e^{i\theta_1}\,\tanh r_1.
\end{equation}

Even though a little tedious, the evaluation of the Gaussian integral in (\ref{eq:over2}) is straightforward enough.
It leads to the following equivalent expressions, by relying on a number of the identities pointed out above.
In one form one finds,
\begin{equation}
\langle\Omega_{z_2}(u_2)|\Omega_{z_1}(u_1)\rangle =
\left(\cosh r_2\cdot\cosh r_1\right)^{-1/2}\,\left(1-\bar{\zeta}_2\zeta_1\right)^{-1/2}\,
e^{\frac{i}{2\hbar}(q_2p_1 - q_1 p_2)}\,e^{-\frac{1}{4}\,G_{(2)}(2,1)},
\end{equation}
where the Gaussian quadratic form $G_{(2)}(2,1)$ is given as
\begin{eqnarray}
&&G_{(2)}(2,1) = \frac{1}{1-\bar{\zeta}_2\zeta_1} \left[
\left(1-\bar{\zeta}_2\right)\left(1-\zeta_1\right)\left(\frac{q_2-q_1}{\ell_0}\right)^2\,-\,\right. \nonumber \\
&-&\left. 2i\left(\bar{\zeta}_2-\zeta_1\right)\left(\frac{q_2-q_1}{\ell_0}\right)\left(\frac{\ell_0}{\hbar}(p_2-p_1)\right)\,+\,
\left(1+\bar{\zeta}_2\right)\left(1+\zeta_1\right)\left(\frac{\ell_0}{\hbar}(p_2-p_1)\right)^2\right].
\end{eqnarray}
In terms of the variables $(u_2,u_1)$ and $(\zeta_2,\zeta_1)$, the same expression writes as,
\begin{equation}
\langle\Omega_{z_2}(u_2)|\Omega_{z_1}(u_1)\rangle =
\left(\cosh r_2\cdot\cosh r_1\right)^{-1/2}\,\left(1-\bar{\zeta}_2\zeta_1\right)^{-1/2}\,
e^{-\frac{1}{2}(u_2\bar{u}_1-\bar{u}_2 u_1)}\,e^{-\frac{1}{4}G_{(2)}(2,1)},
\end{equation}
with this time, in a further streamlined form for the Gaussian quadratic factor,
\begin{eqnarray}
\frac{1}{2}\,G_{(2)}(2,1) &=& \frac{1}{1-\bar{\zeta}_2\zeta_1}
\Bigl((1+\bar{\zeta}_2\zeta_1)|u_2-u_1|^2\,-\,\bar{\zeta}_2(u_2 -u_1)^2\,-\,\zeta_1(\bar{u}_2-\bar{u}_1)^2\Bigr)
\nonumber \\
&=& \frac{1}{1-\bar{\zeta}_2\zeta_1}
\Bigl((u_2-u_1)-\zeta_2(\bar{u}_2-\bar{u}_1)\Bigr)^*
\Bigl((u_2 - u_1)-\zeta_1(\bar{u}_2-\bar{u}_1)\Bigr).
\end{eqnarray}
That the dependence of this result on the different variables parametrising the SR-UR states
$|\Omega_{z_2}(u_2)\rangle$ and $|\Omega_{z_1}(u_1)\rangle$ comes out as established above
may be understood from the following two identities (for the first, see (\ref{eq:Sz})),
\begin{equation}
S(z)|\Omega_0\rangle = (\cosh r)^{-1/2}\,e^{\frac{1}{2}\zeta{a^\dagger}^2}|\Omega_0\rangle,
\end{equation}
\begin{equation}
D^\dagger(u_2)D(u_1)=D(-u_2)D(u_1)=e^{\frac{i}{2\hbar}(q_2p_1-q_1p_2)}\,D(u_1-u_2)
=e^{-\frac{1}{2}(u_2\bar{u}_1 - \bar{u}_2 u_1)}\,D(u_1 - u_2),
\end{equation}
which imply the relation,
\begin{equation}
\langle\Omega_{z_2}(u_2)|\Omega_{z_1}(u_1)\rangle=
\left(\cosh r_2\cdot\cosh r_1\right)^{-1/2}\,e^{\frac{i}{2\hbar}(q_2p_1 - q_1 p_2)}\,
\langle\Omega_0|e^{\frac{1}{2}\bar{\zeta}_2\, a^2}\,D(u_1 - u_2)\,e^{\frac{1}{2}\zeta_1 {a^\dagger}^2}
|\Omega_0\rangle.
\end{equation}
Hence the above evaluations have also established the corresponding general matrix element,
\begin{equation}
\langle\Omega_0|e^{\frac{1}{2}\bar{\zeta}_2\,a^2}\,D(u)\,e^{\frac{1}{2}\zeta_1{a^\dagger}^2}|\Omega_0\rangle=
\left(1-\bar{\zeta}_2\zeta_1\right)^{-1/2}\,{\rm exp}\left\{ -\frac{1}{2}\,\frac{1}{1-\bar{\zeta}_2\zeta_1}
\Bigl(u-\zeta_2\bar{u}\Bigr)^*\Bigl(u-\zeta_1\bar{u})\Bigr)\right\}.
\end{equation}

\section{Conclusions}
\label{Sect7}

In this contribution to the present Workshop Proceedings we have explored the first step in the general programme
outlined in the Introduction, in the case of the quantum observables $Q$ and $P$ defining the Heisenberg
algebra. Namely, generally given a quantum system characterised by a set of quantum observables, one considers
the set of quantum states that saturate the Schr\"odinger-Robertson uncertainty relation corresponding to this
set of observables. Such states are the closest possible to displaying a classical behaviour of the quantum system
while, being determined by a condition characteristic of coherent-like quantum states, they are parametrised
by a collection of continuous variables and therefore define a specific submanifold within the Hilbert space
of the quantum system. Correspondingly there arise specific geometric structures associated to this
manifold, compatible with one another, namely both a quantum symplectic structure as well as
a quantum Riemannian metric. It may even be possible to reconstruct the quantum dynamics of the
system from that geometric data as well as a choice of Hamiltonian operator represented through its
diagonal matrix elements for the saturating states, thereby offering a geometric formulation of quantum systems
and their dynamics through a path integral representation.

This general programme is initiated herein, in the case of observables of the Heisenberg algebra for a single degree of
freedom quantum system as an illustration. Correspondingly the saturating quantum states are the so-called
and well known general squeezed states, for which many properties and results were reviewed and presented
together with quite many details and some original results, with the hope that some readers could become interested
in taking part in such an exploration in the case of other possible choices of quantum observables and
the ensuing saturating quantum states. For instance, the affine quantum algebra of scale transformations,
$[Q,D]=i\hbar\,Q$ (say with $D=(QP+PQ)/2$), does also play an important role in quite many quantum
systems\cite{Klauder1,Klauder2,Klauder3,FZ}, with its own coherent states. To this author's
best knowledge, the analogue states for the affine algebra of the squeezed states for the Heisenberg algebra remain
to be fully understood. Other situations may be thought of as well, such as for instance the operators and uncertainty
relation related to the factorisation of a quantum Hamiltonian along the lines and methods of supersymmetric
quantum mechanics, $H=A^\dagger\,A\,+\,E_0$.

Essential to such a programme is the evaluation of the overlap of the saturating quantum states given a set
of quantum observables. In particular this quantity encodes the data necessary in identifying
the inherent geometric structures, as well as in the construction of the quantum path integral of the system
over the manifold in Hilbert space associated to the saturating quantum states. Usually overcompleteness
relations ensue, implying that the overlap of saturating states determines a reproducing kernel representation
of the Hilbert space.

In this contribution the discussion concludes with the evaluation of this reproducing kernel for the
general squeezed states of the Heisenberg algebra which saturate the Schr\"odinger-Robertson
uncertainty relation. We defer to a separate publication an analysis of the corresponding
symplectic and Riemannian geometric structures, as well as of the path integral representation
of the quantum system over the associated manifold of squeezed states,
$|\Omega_z(u)\rangle=D(u)S(z)|\Omega_0\rangle$. All of these considerations are to follow from the
quantities $\langle\Omega_{z_2}(u_2)|\Omega_{z_1}(u_1)\rangle$.

Thus in particular the overlap $\langle\Omega_{z_2}(u_2)|\Omega_{z_1}(u_1)\rangle$ determines
a reproducing kernel representation of the Hilbert space of the Heisenberg algebra of observables
$Q$ and $P$. Indeed, given the generalised overcompleteness relation (\ref{eq:OC4}),
obviously one has the property,
\begin{equation}
\langle\Omega_{z_2}(u_2)|\Omega_{z_1}(u_1)\rangle=
\int_{\mathbb{C}^2}\frac{du_3\,d\bar{u}_3}{\pi}\,\frac{dz_3\,d\bar{z}_3}{\pi}\,\mu(z_3,\bar{z}_3)\,
\langle\Omega_{z_2}(u_2)|\Omega_{z_3}(u_3)\rangle\,\langle\Omega_{z_3}(u_3)|\Omega_{z_1}(u_1)\rangle.
\end{equation}
We plan to report elsewhere on such applications and further developments
of the results of the present contribution, as well as on the general programme outlined in the
Introduction. This programme is offered here as a token of genuine and sincere appreciation for Professor Norbert
Hounkonnou's constant interest and many scientific contributions of note, and certainly for his unswerving efforts
as well towards the development of mathematical physics in Benin, in Western Africa and on the African continent,
to the benefit of the younger and future generations, and this on the occasion of this special
COPROMAPH Workshop celebrating his sixtieth birthday.

\section*{Acknowledgments}

The author thanks Professor Frederik Scholtz and the National Institute for Theoretical Physics (NITheP, Stellenbosch
node, South Africa) for their warm hospitality during a stay which saw a great part of this work being completed.
This work is supported in part by the Institut Interuniversitaire des Sciences Nucl\'eaires (I.I.S.N., Belgium),
and by the Belgian Federal Office for Scientific, Technical and Cultural Affairs through the Interuniversity
Attraction Poles (IAP) P6/11.

\section*{Appendices}

\subsection*{Appendix A: Cauchy-Schwarz inequality and quantum uncertainty relations}

In the first part of this Appendix, for the purpose of the present paper it proves useful to reconsider
specific arguments leading to the Cauchy-Schwarz inequality. In the second part this inequality is applied
to establish the Schr\"odinger-Robertson uncertainty relation (SR-UR) given any two quantum observables.

Let $|\psi_1\rangle$ and $|\psi_2\rangle$ be any two (normalisable and non-vanishing) quantum states, and consider
their arbitrary complex linear combination, say in the form,
\begin{equation}
|\psi\rangle = |\psi_1\rangle + i \lambda e^{i\varphi} |\psi_2\rangle,\qquad \varphi,\lambda\in\mathbb{R},
\end{equation}
with $\varphi$ a phase factor and $\lambda$ a real parameter. Since the sesquilinear and hermitian inner product
of Hilbert space is positive definite, the norm of the state $|\psi\rangle$ is positive definite whatever the values
for these two parameters,
\begin{equation}
P(\lambda)\equiv\langle\psi|\psi\rangle=
\lambda^2\langle \psi_2|\psi_2\rangle + i\lambda\left(e^{i\varphi}\langle\psi_1|\psi_2\rangle
- e^{-i\varphi}\langle\psi_2|\psi_1\rangle\right) + \langle\psi_1|\psi_1\rangle \ge 0.
\end{equation}
Note that the l.h.s. of this inequality is a real quadratic polynomial in $\lambda\in\mathbb{R}$ with real coefficients,
$P(\lambda)$, of which the coefficient in $\lambda^2$ is strictly positive. Hence the parabolic graph of this function
of $\lambda$ lies entirely in the upper half plane and this polynomial has no real roots in $\lambda$,
unless the state $|\psi\rangle$ itself vanishes identically in which case the two roots are degenerate and real
for just a unique and specific set of values for the parameters $\varphi$ and $\lambda$ such that the parabola
$P(\lambda)$ has its minimum just touching the horizontal coordinate axis in $\lambda$.
Consequently the discriminant of this real quadratic form in $\lambda$ is negative, namely
\begin{equation}
\langle\psi_1|\psi_1\rangle\,\langle\psi_2|\psi_2\rangle \ge -\frac{1}{4}
\left(e^{i\varphi}\langle\psi_1|\psi_2\rangle - e^{-i\varphi}\langle\psi_2|\psi_1\rangle\right)^2 \ge 0.
\label{eq:Discri1}
\end{equation}
This inequality is the tightest when the quantity on the r.h.s. of this relation, which is still a function
of the parameter $\varphi$, reaches its maximal value. As readily established this maximum is
obtained for a phase factor $\varphi=\varphi_0$ such that
\begin{equation}
e^{2i\varphi_0} = -\frac{\langle\psi_2|\psi_1\rangle}{\langle\psi_1|\psi_2\rangle},
\end{equation}
namely,
\begin{equation}
e^{i\varphi_0}\langle\psi_1|\psi_2\rangle=-e^{-i\varphi_0}\langle\psi_2|\psi_1\rangle,\quad
ie^{i\varphi_0}\langle\psi_1|\psi_2\rangle=-ie^{-i\varphi_0}\langle\psi_2|\psi_1\rangle=
\left(ie^{i\varphi_0}\langle\psi_1|\psi_2\rangle\right)^*.
\label{eq:phase0}
\end{equation}
Given this choice, the tightest discriminant inequality in (\ref{eq:Discri1}) reduces to the well-known Cauchy-Schwarz inequality
\begin{equation}
\langle\psi_1|\psi_1\rangle\,\langle\psi_2|\psi_2\rangle \ge |\langle\psi_1|\psi_2\rangle|^2.
\label{eq:Schwarz1}
\end{equation}

Having set the phase factor as $\varphi=\varphi_0$ the polynomial $P(\lambda)$ may be organised in the following form,
\begin{equation}
P(\lambda)=\langle\psi_2|\psi_2\rangle\left[
\left(\lambda+ie^{i\varphi_0}\frac{\langle\psi_1|\psi_2\rangle}{\langle\psi_2|\psi_2\rangle}\right)^2\,+\,
\frac{\langle\psi_1|\psi_1\rangle\langle\psi_2|\psi_2\rangle - |\langle\psi_1|\psi_2\rangle|^2}
{\langle\psi_2|\psi_2\rangle^2}\right]\,\ge\,0.
\end{equation}
Hence making now the additional choice $\lambda=\lambda_{CS}$ such that,
\begin{equation}
\lambda_{CS}=-ie^{i\varphi_0}\frac{\langle\psi_1|\psi_2\rangle}{\langle\psi_2|\psi_2\rangle}=
ie^{-i\varphi_0}\frac{\langle\psi_2|\psi_1\rangle}{\langle\psi_2|\psi_2\rangle},\qquad
ie^{i\varphi_0}\lambda_{CS}=-\frac{\langle\psi_2|\psi_1\rangle}{\langle\psi_2|\psi_2\rangle},
\end{equation}
$\lambda_{CS}$ being indeed a real quantity on account of the properties in (\ref{eq:phase0}),
one has,
\begin{equation}
\langle\psi|\psi\rangle=P(\lambda_{CS})=\langle\psi_2|\psi_2\rangle\,
\frac{\langle\psi_1|\psi_1\rangle\langle\psi_2|\psi_2\rangle - |\langle\psi_1|\psi_2\rangle|^2}
{\langle\psi_2|\psi_2\rangle^2}\,\ge\,0.
\end{equation}
Consequently, besides the Cauchy-Schwarz inequality (\ref{eq:Schwarz1}), one also concludes
that this inequality is saturated into a strict equality provided the states $|\psi_1\rangle$ and $|\psi_2\rangle$
are such that $|\psi\rangle=0$ for these choices of parameters $\varphi=\varphi_0$
and $\lambda=\lambda_{CS}$, namely
\begin{equation}
|\psi_1\rangle\,-\,\frac{\langle\psi_2|\psi_1\rangle}{\langle\psi_2|\psi_2\rangle}\,|\psi_2\rangle=0
\ \Longleftrightarrow\ \langle\psi_1|\psi_1\rangle\,\langle\psi_2|\psi_2\rangle = |\langle\psi_1|\psi_2\rangle|^2.
\label{eq:Schwarz2}
\end{equation}

\vspace{10pt}

Let now $A$ and $B$ be two arbitrary quantum observables, namely hermitian
(and ideally, self-adjoint) operators acting on Hilbert space, $A^\dagger=A$ and $B^\dagger=B$,
and consider an arbitrary (normalisable and non-vanishing) quantum state $|\psi_0\rangle$. Whatever choice
of quantum operator ${\cal O}$, its expectation value for that state $|\psi_0\rangle$ is denoted as
\begin{equation}
\langle{\cal O}\rangle=\frac{\langle\psi_0|{\cal O}|\psi_0\rangle}{\langle\psi_0|\psi_0\rangle}.
\end{equation}
In particular for the observables $A$ and $B$ we have their real valued expectation values
\begin{equation}
a_0=\langle A\rangle,\qquad
b_0=\langle B\rangle,\qquad
a_0, b_0\in\mathbb{R},
\end{equation}
which are used to shift these observables as follows,
\begin{equation}
\bar{A}=A-a_0\mathbb{I},\qquad
\bar{B}=B-b_0\mathbb{I},\qquad
\bar{A}^\dagger=\bar{A},\quad \bar{B}^\dagger=\bar{B},
\end{equation}
such that $[\bar{A},\bar{B}]=[A,B]$. Consequently we have
\begin{equation}
\left(\Delta A\right)^2=\langle\bar{A}^2\rangle,\qquad
\left(\Delta B\right)^2=\langle\bar{B}^2\rangle.
\end{equation}

In order to establish the SR-UR from the Cauchy-Schwarz inequality, let us consider the following
two quantum states,
\begin{equation}
|\psi_1\rangle=\frac{1}{\sqrt{\langle\psi_0|\psi_0\rangle}}\,\bar{A}|\psi_0\rangle,\qquad
|\psi_2\rangle=\frac{1}{\sqrt{\langle\psi_0|\psi_0\rangle}}\,\bar{B}|\psi_0\rangle,
\end{equation}
which are such that
\begin{equation}
\langle\psi_1|\psi_1\rangle=\left(\Delta A\right)^2,\quad
\langle\psi_2|\psi_2\rangle=\left(\Delta B\right)^2,\qquad
\langle\psi_1|\psi_2\rangle=\langle\bar{A}\bar{B}\rangle,\quad
\langle\psi_2|\psi_1\rangle=\langle\bar{B}\bar{A}\rangle=\langle\bar{A}\bar{B}\rangle^*.
\end{equation}
Consequently the Cauchy-Schwarz inequality (\ref{eq:Schwarz1}) reads,
\begin{equation}
\left(\Delta A\right)^2\,\left(\Delta B\right)^2\,\ge |\langle\bar{A}\bar{B}\rangle|^2,\qquad
\left(\Delta A\right)^2\,\left(\Delta B\right)^2\,\ge \langle\bar{A}\bar{B}\rangle\,\langle\bar{B}\bar{A}\rangle.
\end{equation}
Alternatively by expressing the quantity $\langle\bar{A}\bar{B}\rangle$ in terms of the commutator and
anti-commutator of the operators $\bar{A}$ and $\bar{B}$ which then separate its real and imaginary
parts\footnote{Note that $(-i)[\bar{A},\bar{B}]$ and $\left\{\bar{A},\bar{B}\right\}$ are hermitian (or even
self-adjoint if $A$ and $B$ are self-adjoint) operators whose expectations values are thus real.}, namely
\begin{equation}
\langle\bar{A}\bar{B}\rangle=\langle\frac{1}{2}\left[\bar{A},\bar{B}\right]+
\frac{1}{2}\left\{\bar{A},\bar{B}\right\}\rangle=\frac{1}{2}i\,\langle(-i)\left[\bar{A},\bar{B}\right]\rangle
+\frac{1}{2}\langle\left\{\bar{A},\bar{B}\right\}\rangle,
\end{equation}
\begin{equation}
\langle\bar{B}\bar{A}\rangle=\langle\bar{A}\bar{B}\rangle^*=
-\frac{1}{2}i\,\langle(-i)\left[\bar{A},\bar{B}\right]\rangle
+\frac{1}{2}\langle\left\{\bar{A},\bar{B}\right\}\rangle,
\end{equation}
one obtains the inequality in the Schr\"odinger-Robertson form,
\begin{equation}
\left(\Delta A\right)^2\,\left(\Delta B\right)^2\,\ge\,\frac{1}{4}\langle(-i)\left[A,B\right]\rangle^2\,+\,
\frac{1}{4}\langle\left\{\bar{A},\bar{B}\right\}\rangle^2.
\label{eq:SR-URA1}
\end{equation}
As a by-product one also derives the looser generalised Heisenberg uncertainty relation,
\begin{equation}
\left(\Delta A\right)^2\,\left(\Delta B\right)^2\,\ge\,\frac{1}{4}\langle(-i)\left[A,B\right]\rangle^2,\qquad
\left(\Delta A\right)\,\left(\Delta B\right)\ge \frac{1}{2}|\langle(-i)[A,B]\rangle|.
\end{equation}

Furthermore given (\ref{eq:Schwarz2}) the SR-UR (\ref{eq:SR-URA1}) is saturated,
namely $\left(\Delta A\right)^2\left(\Delta B\right)^2=\langle\bar{A}\bar{B}\rangle\langle\bar{B}\bar{A}\rangle$,
provided the state $|\psi_0\rangle$ is such that,
\begin{equation}
\left(\bar{A}-\frac{\langle\bar{B}\bar{A}\rangle}{\left(\Delta B\right)^2}\,\bar{B}\right)\,|\psi_0\rangle=0,\quad
\left(\bar{A}\,-\,\lambda_0\,\bar{B}\right)\,|\psi_0\rangle = 0,\quad
\left(A-\lambda_0 B\right)|\psi_0\rangle=\left(\langle A\rangle - \lambda_0\langle B\rangle\right)|\psi_0\rangle,
\end{equation}
where the complex parameter $\lambda_0$ is given by
$\lambda_0=\langle\bar{B}\bar{A}\rangle/\left(\Delta B\right)^2=\left(\Delta A\right)^2/\langle\bar{A}\bar{B}\rangle$, 
namely
\begin{equation}
\lambda_0=\frac{1}{\left(\Delta B\right)^2}\,\left(\frac{1}{2}\langle\left\{\bar{A},\bar{B}\right\}\rangle\,
-\,\frac{1}{2}i\,\langle (-i)\left[A,B\right]\rangle\right)
=\left(\Delta A\right)^2\,\frac{1}{\left(\frac{1}{2}\langle\left\{\bar{A},\bar{B}\right\}\rangle\,
+\,\frac{1}{2}i\,\langle (-i)\left[A,B\right]\rangle\right)}.
\end{equation}

\subsection*{Appendix B: Baker-Campbell-Hausdorff formulae}

This Appendix is structured in three parts. The first recalls a basic Baker-Campbell-Hausdorff (BCH) formula.
The second part discusses recent results established in Ref.\cite{Visser} based on a construction of the most general
BCH formula which is also outlined. Finally the third part applies these results to a SU(1,1) algebra directly
related to the general squeezed coherent states arising as the states saturating the Schr\"odinger-Robertson
uncertainty relation.

Given any two operators, $A$ and $B$, the following basic BCH is well known\footnote{It suffices to consider
the generating operator in $\lambda$, $e^{\lambda A}Be^{-\lambda A}$, expanded in series in $\lambda$.},
\begin{equation}
e^{A}\,B\,e^{-A}=B+[A,B]+\frac{1}{2!}[A,[A,B]]+\frac{1}{3!}[A,[A,[A,B]]]+\cdots=e^{ad\, A}\,B,
\label{eq:BCH1}
\end{equation}
where the (Lie algebra) adjoint action of the operator $A$ on an operator $X$ is defined by
\begin{equation}
ad\,A\cdot X\equiv [A,X].
\end{equation}
This identify is to be used throughout hereafter. Note that it also implies
\begin{equation}
e^A\,e^B\,e^{-A}=e^{e^{ad\, A}\,B},\qquad
e^A\,e^B=e^{e^{ad\, A}\,B}\,e^A.
\end{equation}
Hence in particular when $[A,B]$ commutes with both $A$ and $B$, we have simply
\begin{equation}
e^A\,e^B=e^{[A,B]}\,e^B\,e^A.
\label{eq:BCH2}
\end{equation}

In order to establish the general BCH formula, first let us consider some operator $A(\lambda)$ function
of a parameter $\lambda$. Then the following identities apply\footnote{As a reminder we have
$\int_0^1dt\,(1-t)^n\,t^m=n!\,m!/(n+m+1)!$ as well as $\int_0^1 dt\,t^n=1/(n+1)$, hence in particular
$\frac{d}{d\lambda}e^{A(\lambda)}=\int_0^1dt\, e^{(1-t)A(\lambda)}\frac{dA(\lambda)}{d\lambda} e^{tA(\lambda)}$.},
\begin{eqnarray}
e^{-A(\lambda)}\,\frac{d}{d\lambda} e^{A(\lambda)} &=&
\int_0^1dt\,e^{-tA(\lambda)}\,\frac{dA(\lambda)}{d\lambda}\,e^{t A(\lambda)} \nonumber \\
&=& \int_0^1 dt\,e^{-t\, ad\, A(\lambda)}\,\frac{dA(\lambda)}{d\lambda} \nonumber \\
&=& \Phi\Bigl(- ad\, A(\lambda)\Bigr)\,\frac{dA(\lambda)}{d\lambda},
\end{eqnarray}
where the function $\Phi(x)$ is given as
\begin{equation}
\Phi(x)=\int_0^1dt\,e^{tx}=\sum_{n=0}^\infty\,\frac{1}{(n+1)!}\,x^n=\frac{e^x -1}{x}.
\end{equation}
This function being such that
\begin{equation}
\Phi(-\ln x)=\frac{x-1}{x\ln x}=\frac{1}{\Psi(x)},
\end{equation}
it proves useful to also introduce the function $\Psi(x)$ defined by\footnote{Note that
$\Psi(e^{-x})=1/\Phi(x)=x/(e^x-1)=\sum_{n=0}^\infty B_n x^n/n!$
is a generating function for Bernoulli numbers, $B_n$.
The author thanks Christian Hagendorf for pointing this out to him.}
\begin{equation}
\Psi(x)=\frac{x\ln x}{x-1}=1+
\sum_{n=1}^\infty\,\frac{(-1)^{n+1}}{n(n+1)}\,x^n,\qquad
\Phi(-\ln x)\,\Psi(x)=1.
\end{equation}

Given two operators $A$ and $B$, the general BCH formula provides an expression for
the operator $C$ defined by
\begin{equation}
C=\ln(e^A\,e^B),\qquad
e^C=e^A\,e^B.
\end{equation}
In order to establish this expression, let us introduce the generating operator $C(\lambda)$ such that
\begin{equation}
e^{C(\lambda)}=e^A\,e^{\lambda\,B},\qquad
C(\lambda)=\ln(e^A\,e^{\lambda\,B}),\qquad C(0)=A.
\end{equation}
The adjoint action of the operator $C(\lambda)$ on any operator $D$ is given as,
\begin{equation}
e^{C(\lambda)}\,D\,e^{-C(\lambda)}=e^A\,e^{\lambda\,B}\,D\,e^{-\lambda\,B}\,e^{-A},\quad
{\rm namely},\quad
e^{ad\,C(\lambda)}\,D=e^{ad\,A}\,e^{\lambda\,ad\,B}\,D,
\end{equation}
which implies,
\begin{equation}
e^{ad\,C(\lambda)}=e^{ad\,A}\,e^{\lambda\,ad\,B},\qquad
ad\,C(\lambda)=\ln(e^{ad\,A}\,e^{\lambda\,ad\,B}).
\end{equation}
On the other hand, since
\begin{equation}
e^{-C(\lambda)}\,\frac{d}{d\lambda}e^{C(\lambda)}
=e^{-\lambda\,B}\,e^{-A}\,\frac{d}{d\lambda}\,e^A\,e^{\lambda B},\quad {\rm namely},\quad
\Phi(-ad\,C(\lambda))\,\frac{dC(\lambda)}{d\lambda}=B,
\end{equation}
necessarily
\begin{equation}
\frac{dC(\lambda)}{d\lambda}=\Psi\Bigl(e^{ad\,A}\,e^{\lambda\,ad\,B}\Bigr)\,B.
\end{equation}
Given the integration condition $C(0)=A$, finally the following general BCH formulae applies for the operator $C$,
\begin{eqnarray}
C &=& \ln(e^A\,e^B)=A+\int_0^1d\lambda\,\Psi\Bigl(e^{ad\,A}\,e^{\lambda\,ad\,B}\Bigr)\,B \nonumber \\
&=&A\,+\,B\,-\,\int_0^1d\lambda\,\sum_{n=1}^\infty\,\frac{1}{n(n+1)}\,\Bigl(\mathbb{I}\,-\,
e^{ad\,A}\,e^{\lambda\,ad\,B}\Bigr)^n\,B \nonumber \\
&=&A\,+\,B\,+\,\int_0^1d\lambda\,\sum_{n=1}^\infty\,\frac{1}{n(n+1)}\,
\Bigl(\mathbb{I}\,-\,e^{ad\,A}\,e^{\lambda\,ad\,B}\Bigr)^{n-1}\,
\left(\frac{e^{ad\,A}-\mathbb{I}}{ad\,A}\right)\,[A,B].
\label{eq:BCHgen}
\end{eqnarray}
In particular when $[A,B]$ commutes with both $A$ and $B$, one has the well known BCH formula,
\begin{equation}
C=\ln(e^A\,e^B)=A+B+\frac{1}{2}[A,B],\qquad
e^A\,e^B=e^{A+B+\frac{1}{2}[A,B]}=e^{\frac{1}{2}[A,B]}\,e^{A+B},
\label{eq:BCH3}
\end{equation}
which implies again the result in (\ref{eq:BCH2}).

It is the last form for the BCH formula in (\ref{eq:BCHgen}) which is the starting point of the recent analysis
of Ref.\cite{Visser} which manages to sum up the BCH formula in closed form in the case of operators
$A$ and $B$ whose commutator is of the form,
\begin{equation}
[A,B]=uA\,+\,vB\,+\,c\mathbb{I},
\end{equation}
where $u$, $v$ and $c$ are constant parameters\footnote{Note that for all practical purposes
the results of Ref.\cite{Visser} remain valid as stated if the term $c\mathbb{I}$ stands for an
operator which commutes with both $A$ and $B$.}. Indeed in such a situation one has,
\begin{eqnarray}
ad\,A\cdot[A,B] &=& v\,[A,B],\quad
ad\,B\cdot[A,B]=-u\,[A,B], \nonumber \\
e^{ad\,A}\,[A,B] &=& e^v\,[A,B],\quad
e^{\lambda\,ad\,B}\,[A,B]=e^{-\lambda u}\,[A,B],
\end{eqnarray}
which implies that
\begin{equation}
\ln(e^A\,e^B)=A\,+\,B\,+f(u,v)\,[A,B],
\label{eq:Vis1}
\end{equation}
where $f(u,v)$ is a simple function determined from (\ref{eq:BCHgen}) in the form
\begin{equation}
f(u,v)=\frac{e^v-1}{v}\,\int_0^1d\lambda\,\sum_{n=1}^\infty\,\frac{1}{n(n+1)}\,
\left(1-e^v\,e^{-\lambda\,v}\right)^{n-1}.
\end{equation}
A direct evaluation finds for this function, which proves to be symmetric,
\begin{equation}
f(u,v)=\frac{u e^u(e^v-1)-v e^v(e^u-1)}{uv(e^u-e^v)}=
\frac{u(1-e^{-v})-v(1-e^{-u})}{uv(e^{-v}-e^{-u})}=f(v,u),
\end{equation}
with the distinguished value $f(0,0)=1/2$ (in agreement with (\ref{eq:BCH3})).

Finally given the reference Fock algebra of operators $a$ and $a^\dagger$ introduced in Section \ref{Sect3},
consider the operators
\begin{equation}
K_0=\frac{1}{2}\left(a^\dagger a+\frac{1}{2}\right),\quad
K_+=\frac{1}{2}\,{a^\dagger}^2,\quad
K_-=\frac{1}{2}\,a^2,
\end{equation}
which generate a SU(1,1) algebra of transformations acting on the Hilbert space representing the Heisenberg
algebra of observables $Q$ and $P$,
\begin{equation}
[K_0,K_\pm]=\pm\,K_\pm,\qquad
[K_-,K_+]=2K_0.
\end{equation}
Independently of the representation which realises this SU(1,1) algebra, let us apply the result (\ref{eq:Vis1})
to specific combinations of operators $K_0$ and $K_\pm$ obeying this algebraic structure.

To begin with consider the following operator
\begin{equation}
e^{\alpha\, K_+}\,e^{\gamma\,K_0}\,e^{-\bar{\alpha}\,K_-},
\label{eq:comb1}
\end{equation}
where $\alpha$ is an arbitrary complex parameter such that $|\alpha|<1$ and $\gamma=\ln(1-|\alpha|^2)$.
In order to apply (\ref{eq:Vis1}), let us rewrite this operator in the form\cite{Matone},
\begin{equation}
e^{\alpha\, K_+}\,e^{\gamma\,K_0}\,e^{-\bar{\alpha}\,K_-}=
e^{\alpha\,K_+}\,e^{\gamma_s\,K_0}\,e^{\gamma_{-s}\,K_0}\,e^{-\bar{\alpha}\,K_-},
\label{eq:split}
\end{equation}
where
\begin{equation}
\gamma_s=\ln(1+s |\alpha|),\quad
\gamma_s\,+\,\gamma_{-s}=\gamma_+\,+\,\gamma_- = \gamma=\ln(1-|\alpha|^2),\quad
s=\pm\,1.
\end{equation}
Since we have
\begin{equation}
[\alpha\,K_+,\gamma_s\,K_0]=-\gamma_s\,(\alpha\,K_+),\qquad
[\gamma_{-s}\,K_0,-\bar{\alpha}\,K_-]=-\gamma_{-s}\,(-\bar{\alpha}\,K_-),
\end{equation}
the BCH formula (\ref{eq:Vis1}) then applies separately to the first two, and the last two factors
in (\ref{eq:split}). With the evaluation of the corresponding values for the function $f(u,v)$, one then finds,
\begin{eqnarray}
\ln(e^{\alpha\,K_+}\,e^{\gamma_s\,K_0}) &=& \frac{s}{|\alpha|}\gamma_s\,\alpha\,K_+\,+\,\gamma_s\,K_0\equiv \tilde{A},
\nonumber \\
\ln(e^{\gamma_{-s}\,K_0}\,e^{-\bar{\alpha}\,K_-}) &=&
\frac{s}{|\alpha|}\gamma_{-s}\,\bar{\alpha}\,K_-\,+\,\gamma_{-s}\,K_0\equiv \tilde{B},
\end{eqnarray}
namely so far,
\begin{equation}
e^{\alpha\, K_+}\,e^{\gamma\,K_0}\,e^{-\bar{\alpha}\,K_-}=e^{\tilde{A}}\,e^{\tilde{B}}.
\end{equation}
However the values for $\gamma_s$ are chosen not only such that $\gamma_s+\gamma_{-s}=\gamma$
but also such that the BCH formula (\ref{eq:Vis1}) may be applied once again to the latter product, which
requires that the commutator of $\tilde{A}$ and $\tilde{B}$ be again a linear combination of these same two operators,
\begin{equation}
[\tilde{A},\tilde{B}]=-\gamma_{-s}\,\tilde{A}\,-\,\gamma_s\,\tilde{B}.
\end{equation}
The final evaluation of the BCH formula for (\ref{eq:comb1}) then reduces to the determination of the
value $f(-\gamma_{-s},-\gamma_s)$. In order to present this BCH formula, since $|\alpha|<1$
let us introduce the following parameters, with $0\le r<\infty$ and $-\pi\le\theta<+\pi$,
\begin{equation}
|\alpha|=\tanh r,\qquad
\alpha=e^{i\theta}\,\tanh r,\qquad
z=e^{i\theta}\,r.
\end{equation}
One then has finally,
\begin{equation}
e^{\alpha\,K_+}\,e^{\ln(1-\tanh^2 r)\,K_0}\,e^{-\bar{\alpha}\,K_-}=e^{z\,K_+\,-\,\bar{z}\,K_-}.
\label{eq:BCH-Sz}
\end{equation}
A similar procedure may be applied to the operator
\begin{equation}
e^{-\bar{\alpha}\,K_-}\,e^{-\gamma\,K_0}\,e^{\alpha \,K_+}.
\end{equation}
However given the following inner automorphism of the SU(1,1) algebra,
\begin{equation}
K_0 \longleftrightarrow -K_0,\qquad
K_+ \longleftrightarrow - K_-,\qquad
K_- \longleftrightarrow - K_+,
\end{equation}
from (\ref{eq:BCH-Sz}) one readily has (with then also $\alpha\leftrightarrow\bar{\alpha}$ and
$z\leftrightarrow\bar{z}$),
\begin{equation}
e^{-\bar{\alpha}\,K_-}\,e^{-(1-\tanh^2 r)\,K_0}\,e^{\alpha \,K_+}=e^{z\,K_+\,-\,\bar{z}\,K_-}=S(z)
=e^{\alpha\,K_+}\,e^{\ln(1-\tanh^2 r)\,K_0}\,e^{-\bar{\alpha}\,K_-}.
\label{eq:BCH-Sz2}
\end{equation}
The results (\ref{eq:BCH-Sz}) and (\ref{eq:BCH-Sz2}), which thus apply to the squeezing operator $S(z)$
introduced in Section~\ref{Subsect4.3}, are stated in Ref.\cite{Pere}
by establishing them in the defining representation of the SU(1,1) algebra.
Here they are derived solely from the structure of the SU(1,1) algebra and independently
of the representation of that algebra, by using the conclusions of Ref.\cite{Visser}.

\end{document}